\documentclass[journal]{IEEEtran}

\IEEEoverridecommandlockouts

\usepackage[cmex10]{amsmath}
\usepackage{amssymb}
\usepackage{mdframed}
\usepackage{cases}
\usepackage{multirow}
\usepackage{empheq}

\usepackage{siunitx}

\usepackage[noadjust]{cite}
\usepackage{verbatim}

\usepackage{algorithm}
\usepackage{algpseudocode}

\ifCLASSINFOpdf
\usepackage[pdftex]{graphicx}
\else
\usepackage[dvips]{graphicx}
\fi
\usepackage[caption=false,font=footnotesize]{subfig}

\usepackage{color}
\usepackage{multirow}

\usepackage[hyphens]{url}

\newtheorem{proposition}{Proposition}

\begin{document}
	
    \sloppy
	
    \title{Robust Deep Joint Source Channel Coding for Task-Oriented Semantic Communications}

    \author{
	\IEEEauthorblockN{Taewoo Park, Eunhye Hong, Yo-Seb Jeon, Namyoon Lee, and Yongjune Kim}	\\		
    
    \thanks{

        Taewoo Park, Eunhye Hong, Yo-Seb Jeon, Namyoon Lee, and Yongjune Kim are with the Department of Electrical Engineering, Pohang University of Science and Technology (POSTECH), Pohang 37673, South Korea (e-mail: \{parktaewoo, eunhye.hong, yoseb.jeon, nylee, yongjune\}@postech.ac.kr).
          
        }		
    }
	
	
    \maketitle
    	
    \begin{abstract}
    Semantic communications based on deep joint source-channel coding (JSCC) aim to improve communication efficiency by transmitting only task-relevant information.
    However, ensuring robustness to the stochasticity of communication channels remains a key challenge in learning-based JSCC.
    In this paper, we propose a novel regularization technique for learning-based JSCC to enhance robustness against channel noise.
    The proposed method utilizes the Kullback-Leibler (KL) divergence as a regularizer term in the training loss, measuring the discrepancy between two posterior distributions: one under noisy channel conditions (noisy posterior) and one for a noise-free system (noise-free posterior).
    Reducing this KL divergence mitigates the impact of channel noise on task performance by keeping the noisy posterior close to the noise-free posterior.  
    We further show that the expectation of the KL divergence given the encoded representation can be analytically approximated using the Fisher information matrix and the covariance matrix of the channel noise.
    Notably, the proposed regularization is architecture-agnostic, making it broadly applicable to general semantic communication systems over noisy channels. 
    Our experimental results validate that the proposed regularization consistently improves task performance across diverse semantic communication systems and channel conditions.

    \end{abstract}  

    
    \section{Introduction}
           
    Semantic communications~\cite{Gunduz2022beyond} are emerging as a prominent research area in next-generation communication systems~\cite{Shi2023task, Hoydis2021toward}.
    The advanced applications envisioned for 6G networks such as autonomous driving, drones, augmented reality (AR), and virtual reality (VR) impose ultra-reliable and low-latency communications (xURLLC) requirements~\cite{Hong20226g}.
    To achieve these stringent demands, recent advancements in deep learning have driven the development of semantic communication systems that extract and transmit the desired meaning of the data rather than prioritizing symbol-level accuracy. 
    Unlike traditional communication systems that aim to accurately transmit symbols or bit sequences, semantic and task-oriented communications focus on transmitting only the information critical for the given task.
    This approach enhances communication efficiency by ensuring the required task performance without fully reconstructing the original data.
    Numerous studies have shown that semantic communications can outperform traditional methods across a range of applications~\cite{Xie2021deep, Tung2022deepwive, Weng2021semantic,Kim2023distributed, Li2023domain, Huang2022toward, Im2024attention}.
    
    As an effective method to realize semantic communications, deep learning-based joint source-channel coding (JSCC) has gained significant attention~\cite{Bourtsoulatze2019deep, Choi2019neural}.
    This approach leverages neural networks to jointly model and optimize source coding and channel coding.
    Within the framework of learning-based JSCC, semantic and task-oriented communications can be implemented through a unified approach that integrates compression, transmission, and task execution into a single neural network system, enabling joint optimization via the backpropagation algorithm.
    The effectiveness of this approach has been demonstrated across a wide range of tasks~\cite{Zhang2023adaptive, Erdemir2023generative, Kurka2020deep, Farsad2018deep, Kondi2001joint}.
    
    Despite its advantages, a key challenge in deep learning-based JSCC lies in addressing the stochastic nature of communication channels~\cite{Bourtsoulatze2019deep,Shao2022learning, Xie2023robust}.
    Designing semantic communication systems that can produce reliable outputs under varying channel noise is crucial to overcoming this challenge.
    A common approach is end-to-end training, where the channel is modeled as a non-trainable layer inserted between the encoder and decoder~\cite{Bourtsoulatze2019deep}. 
    While this method improves robustness compared to training without accounting for noise, it remains susceptible to channel variations. 
    When the model parameters are trained for a specific channel environment, its task performance can degrade significantly if the actual channel conditions differ from the trained environment.
    
    To enhance the robustness to channel variations, several approaches have been proposed.
    Shao~\mbox{\textit{et al.}}~\cite{Shao2022learning} developed a variable-length variational feature encoding (VL-VFE) method that dynamically adjusts the number of transmitted symbols according to the channel noise variance.
    Xie~\mbox{\textit{et al.}}~\cite{Xie2023robust} introduced a robust information bottleneck (RIB) principle, formulating a trade-off between task-relevant information and redundancy.
    Park~\mbox{\textit{et al.}}~\cite{Park2025joint} proposed a JSCC approach for channel-adaptive digital semantic communication. 
    These prior works regulate metrics such as activated dimensions~\cite{Shao2022learning}, encoded redundancy~\cite{Xie2023robust}, and modulation/demodulation orders~\cite{Park2025joint}.
    In contrast, we focus on improving the robustness of the posterior distribution, which directly impacts task performance. 
    For instance, in a classification task, the final inference result corresponds to the class that maximizes the posterior distribution estimated by the receiver's classifier. 
    To our knowledge, deep learning-based JSCC schemes that explicitly improve the robustness of the posterior distribution have not yet been explored.     
    
    \begin{figure*}[t] 
        \centering
        \includegraphics[width=0.65\textwidth]{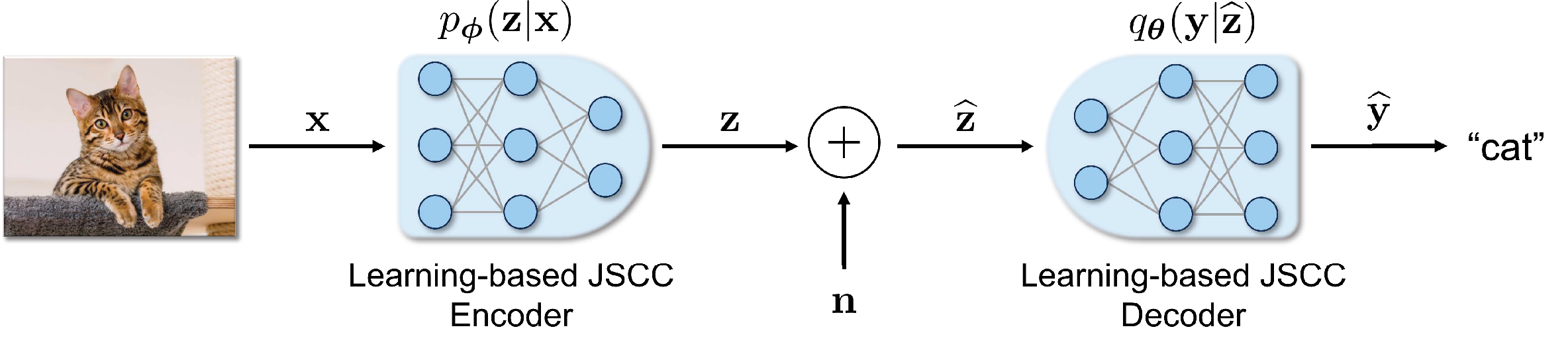}
        \caption{The system model of learning-based JSCC schemes for classification tasks. The input image data is denoted as $\mathbf{x}$, the encoded representation as $\mathbf{z}$, the channel noise as $\mathbf{n}$, the noisy received representation as $\widehat{\mathbf{z}}$, and the estimated label as $\widehat{\mathbf{y}}$. The encoder maps the image $\mathbf{x}$ to the representation $\mathbf{z}$ based on the conditional probability distribution $p_{\boldsymbol{\phi}}(\mathbf{z}|\mathbf{x})$, while the decoder estimates the label $\widehat{\mathbf{y}}$ from the received representation $\widehat{\mathbf{z}}$ following $q_{\boldsymbol{\theta}}(\mathbf y|\widehat{\mathbf{z}})$. Both $p_{\boldsymbol{\phi}}(\mathbf{z}|\mathbf{x})$ and $q_{\boldsymbol{\theta}}(\mathbf y|\widehat{\mathbf{z}})$ are parametrized by neural networks, which function as the encoder and decoder, respectively.}
        \label{fig:system model}
        \vspace{-3mm}
    \end{figure*}

    In this paper, we propose a novel regularization technique for learning-based JSCC to enhance robustness against channel noise. 
    In semantic and task-oriented communications, robustness refers to maintaining reliable task performance regardless of channel noise. 
    This can be achieved by ensuring consistency in the posterior distribution for a given task, even when channel noise perturbs the encoded representations.
    Hence, we aim to minimize the discrepancy between two posterior distributions: one under noisy channel conditions (noisy posterior) and one for a noise-free system (noiseless posterior).
    We introduce a regularization technique based on the expected Kullback-Leibler (KL) divergence, which quantifies the average distance between these posteriors. 
    By incorporating this regularization term into the training loss, the model parameters are optimized to reduce the impact of channel noise by ensuring the noisy posterior remains close to the noise-free posterior. 


    Furthermore, we show that the regularization term based on the KL divergence can be analytically expressed using the Fisher information matrix and the covariance matrix of the channel noise.
    In particular, for an additive white Gaussian noise (AWGN) channel, the regularization term can be derived as the product of the trace of the Fisher information matrix and the noise variance.
    Since the Fisher information matrix corresponds to the expectation of the Hessian of negative log posterior, minimizing this term smooths the curvature of the log posterior, thereby enhancing robustness against channel noise. 
    In addition, the degree of smoothing adapts to the channel condition, as the regularization term explicitly incorporates the noise variance.
    Notably, the proposed regularization is architecture-agnostic and broadly applicable to general semantic communication systems since it requires only a differentiable posterior. 
    Importantly, the proposed method does not increase computational complexity during inference, as the modifications are limited to the loss function during training.

    To evaluate the effectiveness of the proposed regularization, we apply it to several recent leaning-based JSCC schemes for classification tasks.  
    The experimental results show that the proposed regularization consistently improves classification accuracy in both analog and digital semantic communication settings, as well as under both fixed and varying channel conditions during training. 
    In addition, we present a visualization of how the proposed regularization affects the curvature of the posterior distribution, providing deeper insights into its effectiveness. 
    
    Our main contributions are as follows:
    \begin{enumerate}
        \item We propose a novel regularization technique based on the KL divergence, which quantifies the discrepancy between the noisy posterior and the noise-free posterior. 
        Specifically, we derive that the expectation of this KL divergence conditioned on the encoded representation, can be analytically approximated using the Fisher information matrix and the covariance matrix of the channel noise. 
        \item We demonstrate that the proposed regularization technique smooths the curvature of the log posterior distribution.
        Furthermore, we show that the degree of smoothing adapts dynamically to the channel condition, as the regularization term incorporates the noise variance.
        \item We validate the effectiveness of the proposed regularization through extensive experiments on several learning-based JSCC schemes for classification tasks. 
        The results indicate that the proposed regularization consistently improves classification accuracy across diverse scenarios.
    \end{enumerate}
    
    
    The rest of this paper is organized as follows. Section \ref{sec:system_model} introduces the system model. Section \ref{sec:main} presents the derivation of the proposed regularization and its theoretical implications. Section \ref{sec:results} provides experimental results and Section \ref{sec:conclusion} concludes the paper.
    
    \section{System Model}\label{sec:system_model}
    
    In this paper, we focus on task-oriented communication using the learning-based JSCC method. 
    The system model for the considered learning-based JSCC scheme applied to a classification task is shown in Fig.~\ref{fig:system model}.
    The task-oriented communication can be represented by the following Markov chain~\cite{Shao2022learning}: 
    \begin{equation}\label{system model}
        Y \to X\to Z\to \widehat{Z} \to \widehat{Y}.
    \end{equation}
    Here, the input data (e.g., image) and the target data (e.g., label) are defined as a pair of random variables $(X, Y)$ that follow the joint distribution $p(\mathbf {x},\mathbf{y})$. 
    The transmitter encodes the input data $\mathbf{x}\in \mathbb{R}^{m}$ into a representation $\mathbf{z}\in \mathbb{R}^{k}$ according to the conditional probability distribution $p(\mathbf{z}|\mathbf{x})$. 
    The encoded representation $\mathbf{z}$ is then transmitted over a noisy channel defined by the transition probability $p(\widehat{\mathbf{z}}|\mathbf{z})$. 
    We consider an AWGN channel, where the received output $\widehat{\mathbf{z}}$ is given by \begin{equation}\widehat{\mathbf{z}}=\mathbf{z}+\mathbf{n},\end{equation}where the noise vector is denoted as $\mathbf{n}\sim \mathcal{N}(0,\sigma^2 I)$. 
    Here, $I$ denotes the identity matrix. 
    The objective of the receiver is to estimate the target variable using the posterior distribution $p(\mathbf{y}|\widehat{\mathbf{z}})$. 
    
    In the context of learning-based JSCC, the probability distribution for encoding $p_{\boldsymbol{\phi}}(\mathbf{z}|\mathbf{x})$ is parameterized by a neural network with parameters $\boldsymbol{\phi}$. 
    We employ a deterministic encoder $\mathbf{z}=f_{\boldsymbol{\phi}}(\mathbf{x})$, where the distribution can be regarded as a Dirac-delta function. 
    The posterior distribution for decoding is modeled by $q_{\boldsymbol{\theta}}(\mathbf{y}|\widehat{\mathbf{z}})$, where $\boldsymbol{\theta}$ denotes the parameters of the neural network at the receiver.
    The objective of learning-based JSCC is to jointly optimize the encoder parameters $\boldsymbol{\phi}$ and the decoder parameters $\boldsymbol{\theta}$ to maximize both task performance and communication efficiency. 

    \section{Robust Learning-Based JSCC}\label{sec:main}
    
    In this section, we propose a novel regularization technique to enhance the robustness of learning-based JSCC, derive its formulation, and discuss its implications.
    
    \subsection{Problem Description}
    
    In task-oriented communications, robustness implies that task performance should remain consistent regardless of channel noise.
    To achieve this consistency, the impact of channel noise on the posterior distribution should be minimized. 
    Specifically, the posterior distribution without channel noise $q_{\boldsymbol{\theta}}(\mathbf{y}|\mathbf{z})$ (i.e., noise-free posterior) and the posterior distribution with channel noise $q_{\boldsymbol{\theta}}(\mathbf{y}|\widehat{\mathbf{z}})$ (i.e., noisy posterior) should be close. 
    The discrepancy between these distributions can be quantified using the KL divergence as follows:  
    \begin{equation}\label{eq:KLdiv}D_{KL}\left(q_{\boldsymbol{\theta}}(\mathbf{y}|\mathbf{z})\middle\|q_{\boldsymbol{\theta}}(\mathbf{y}|\widehat{\mathbf{z}})\right)=\sum_{\mathbf{y}\in\mathcal{Y}}q_{\boldsymbol{\theta}}(\mathbf y|\mathbf{z})\log \frac{q_{\boldsymbol{\theta}}(\mathbf y|\mathbf{z})}{q_{\boldsymbol{\theta}}(\mathbf y|\mathbf{\widehat z})}.
    \end{equation}
    The KL divergence in \eqref{eq:KLdiv} quantifies the difference between two probability mass functions of $\mathbf y$ given the noise-free representation $\mathbf{z}$ and the noisy representation $\widehat{\mathbf{z}}$. 
    
    This KL divergence differs slightly  from the standard definition of conditional KL divergence in~\cite{Cover2006elements}, which is typically defined with the same conditioning variable and measures the expected KL divergence over that variable.
        
    To compute the expected KL divergence between these distributions across all $\mathbf{z}$ and $\widehat{\mathbf{z}}$, we take the following expectation:    
    \begin{align}
    &\mathbb{E}_{p_{\boldsymbol{\phi}}(\mathbf z,\widehat{\mathbf z})}\left[D_{KL}\left(q_{\boldsymbol{\theta}}(\mathbf{y}|\mathbf{z})\middle\|q_{\boldsymbol{\theta}}(\mathbf{y}|\widehat{\mathbf{z}})\right)\right] \nonumber \\
    &=\mathbb{E}_{p_{\boldsymbol{\phi}}(\mathbf z)}\left[\mathbb{E}_{p(\widehat{\mathbf z}|\mathbf z)}\left[D_{KL}\left(q_{\boldsymbol{\theta}}(\mathbf{y}|\mathbf{z})\middle\|q_{\boldsymbol{\theta}}(\mathbf{y}|\widehat{\mathbf{z}})\right)\right]\right] \nonumber \\
    &=\mathbb{E}_{p(\mathbf x)}\left[\mathbb{E}_{p(\widehat{\mathbf z}|\mathbf z=f_{\boldsymbol{\phi}}(\mathbf{x}))}\left[D_{KL}\left(q_{\boldsymbol{\theta}}(\mathbf{y}|\mathbf{z})\middle\|q_{\boldsymbol{\theta}}(\mathbf{y}|\widehat{\mathbf{z}})\right)\right]\right],\label{eq:expectedKLdiv3}
    \end{align}
    where~\eqref{eq:expectedKLdiv3} follows from the deterministic nature of the encoding function.
    We propose using the expected KL divergence as a regularization term to enhance robustness. 
    By minimizing this regularization term, we aim to reduce the average discrepancy between the noisy posterior and the noise-free posterior. 
    Since the decoder relies on the posterior distribution to perform tasks, minimizing~\eqref{eq:expectedKLdiv3} improves the robustness of task performance against channel noise.
    Moreover, under the AWGN channel model, minimizing the proposed regularization can be interpreted as reducing the discrepancy between the posterior distribution conditioned on the noisy representation $\widehat{\mathbf{z}}$ and the posterior distribution conditioned on its minimum mean squared error (MMSE) estimate, $\mathbf{z}=\mathbb{E}\left[\widehat{\mathbf{z}}\right]$.
    
    \subsection{Regularization for Robustness}
    
    We derive an analytical approximation of the proposed regularization term~\eqref{eq:expectedKLdiv3} and describe the smoothing effect induced by this regularization.
    In~\eqref{eq:expectedKLdiv3}, the outer expectation depends only on the source data distribution and can be computed by sampling data points from this distribution. 
    The inner expectation pertains to the channel and can be computed analytically using a Taylor approximation.
    The second order Taylor approximation of the KL divergence $D_{KL}\left(q_{\boldsymbol{\theta}}(\mathbf{y}|\mathbf{z})\middle\|q_{\boldsymbol{\theta}}(\mathbf{y}|\widehat{\mathbf{z}})\right)$ is given in the following proposition.
    \begin{proposition}\label{prop:KL_Taylor}
    By treating the KL divergence~\eqref{eq:KLdiv} as a function of $\widehat{\mathbf{z}}$ and applying the second order Taylor approximation around $\mathbf{z}$, the KL divergence can be approximated as
    \begin{equation} \label{eq:Taylor}
    D_{KL}\left(q_{\boldsymbol{\theta}}(\mathbf{y}|\mathbf{z})\middle\|q_{\boldsymbol{\theta}}(\mathbf{y}|\widehat{\mathbf{z}})\right)
    \approx \frac{1}{2}(\widehat{\mathbf{z}}-\mathbf{z})^{\mathsf T}\mathcal{I}(\mathbf{z})(\widehat{\mathbf{z}}-\mathbf{z}),
    \end{equation}
    where $\mathcal{I}(\mathbf{z})$ is the Fisher information matrix defined as 
    \begin{equation} \label{eq:fisher}
        \mathcal{I}(\mathbf{z})=\mathbb{E}_{q_{\boldsymbol{\theta}}(\mathbf y|\mathbf{z})}\left[\nabla_{\mathbf z} \log q_{\boldsymbol{\theta}}(\mathbf y|\mathbf{z})\nabla_{\mathbf z} \log q_{\boldsymbol{\theta}}(\mathbf y|\mathbf{z})^{\mathsf T}\right].
    \end{equation}
    \end{proposition}
    \begin{IEEEproof}
        See Appendix~\ref{sec:proof1}.
    \end{IEEEproof}
    
    Proposition~\ref{prop:KL_Taylor} shows that the KL divergence can be approximated using the Fisher information matrix of the encoded representation $\mathbf z$. 
    Furthermore, the second order approximation of the KL divergence can be efficiently computed using only the gradient of the log posterior as shown in \eqref{eq:fisher}. 
    
    Based on the Taylor approximation in Proposition~\ref{prop:KL_Taylor}, we derive the conditional expectation of the KL divergence with respect to the channel transition probability $p(\widehat{\mathbf{z}}|\mathbf{z})$. 

    \begin{proposition} \label{prop:expected_KL}
    The conditional expectation of the KL divergence $D_{KL}\left(q_{\boldsymbol{\theta}}(\mathbf{y}|\mathbf{z})\middle\|q_{\boldsymbol{\theta}}(\mathbf{y}|\widehat{\mathbf{z}})\right)$ given the encoded representation $\mathbf{z}$ can be approximated as 
    \begin{equation}
        \mathbb{E}_{p(\widehat{\mathbf z}|\mathbf z)}\left[ D_{KL}\left(q_{\boldsymbol{\theta}}(\mathbf{y}|\mathbf{z})\middle\|q_{\boldsymbol{\theta}}(\mathbf{y}|\widehat{\mathbf{z}})\right)\right] \approx \frac{1}{2}\operatorname{Tr}\left(\mathcal{I}(\mathbf{z}) \cdot \Sigma_\mathbf{n}\right), \label{eq:regularizer_main}
    \end{equation}
    where the $\Sigma_\mathbf{n}$ is the covariance matrix of noise $\mathbf{n}$. 
    For the AWGN channel, the conditional expectation can be approximated as
    \begin{equation}
    \mathbb{E}_{p(\widehat{\mathbf z}|\mathbf z)}\left[ D_{KL}\left(q_{\boldsymbol{\theta}}(\mathbf{y}|\mathbf{z})\middle\|q_{\boldsymbol{\theta}}(\mathbf{y}|\widehat{\mathbf{z}})\right)\right] \approx 
    \frac{\sigma^2}{2}\operatorname{Tr}\left(\mathcal{I}(\mathbf{z})\right),
    \end{equation}
    where the remainder term becomes negligible as $\sigma\to 0$.
    For the Rayleigh slow fading channel with a given $h$, where $\widehat{\mathbf{z}}=\mathbf{z}+\mathbf{n}/|h|$, the conditional expectation can be approximated as
    \begin{equation}
    \mathbb{E}_{p(\widehat{\mathbf z}|\mathbf z)}\left[ D_{KL}\left(q_{\boldsymbol{\theta}}(\mathbf{y}|\mathbf{z})\middle\|q_{\boldsymbol{\theta}}(\mathbf{y}|\widehat{\mathbf{z}})\right)\right] \approx 
    \frac{\sigma^2}{2|h|^2}\operatorname{Tr}\left(\mathcal{I}(\mathbf{z})\right).
    \end{equation}    
    \end{proposition}    
    \begin{IEEEproof}
        See Appendix~\ref{sec:proof2}.
    \end{IEEEproof}
    
    Proposition~\ref{prop:expected_KL} indicates that the conditional expectation of the KL divergence can be computed approximately if the channel's noise covariance matrix is known, including both AWGN and colored Gaussian noise channels.
    For both AWGN and Rayleigh fading channels, the regularization terms share the same form of $c\operatorname{Tr}(\mathcal{I}(\mathbf{z}))$, where $c$ is a constant.
    Thus, we set the regularization term $\mathcal{R}(\boldsymbol{\phi},\boldsymbol{\theta},\mathbf{z})$ as
    \begin{equation}\label{eq:cond_reg}
        \mathcal{R}(\boldsymbol{\phi},\boldsymbol{\theta},\mathbf{z}) \triangleq\frac{\sigma^2}{2}\operatorname{Tr}(\mathcal{I}(\mathbf{z})).
    \end{equation}
    This regularization term $\mathcal{R}(\boldsymbol{\phi},\boldsymbol{\theta},\mathbf{z})$ quantifies how much the noisy posterior can deviate from the noise-free posterior, given the encoded representation $\mathbf{z}$. 
    Moreover, under deterministic encoding, it becomes $\mathcal{R}\left(\boldsymbol{\phi},\boldsymbol{\theta},f_{\boldsymbol{\phi}}(\mathbf{x})\right)$, which measures the expected difference between posteriors conditioned on the input data $\mathbf{x}$. 

    \begin{figure}[t] 
        \centering
        \subfloat[][Good channel condition]{\includegraphics[width=0.24\textwidth]{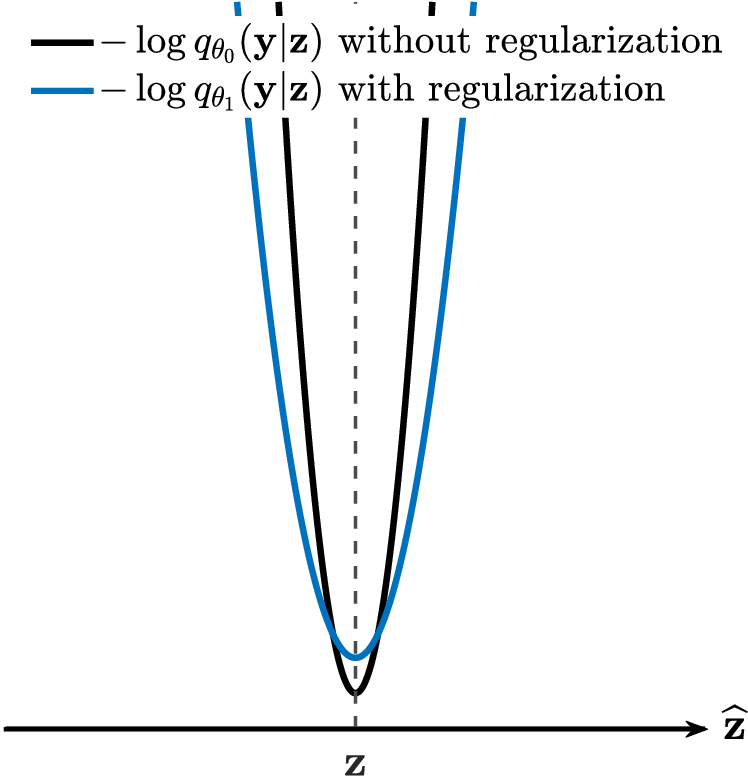}\label{fig:curvature_good}}
        \hspace{0.02cm}
        \subfloat[][Poor channel condition]{\includegraphics[width=0.24\textwidth]{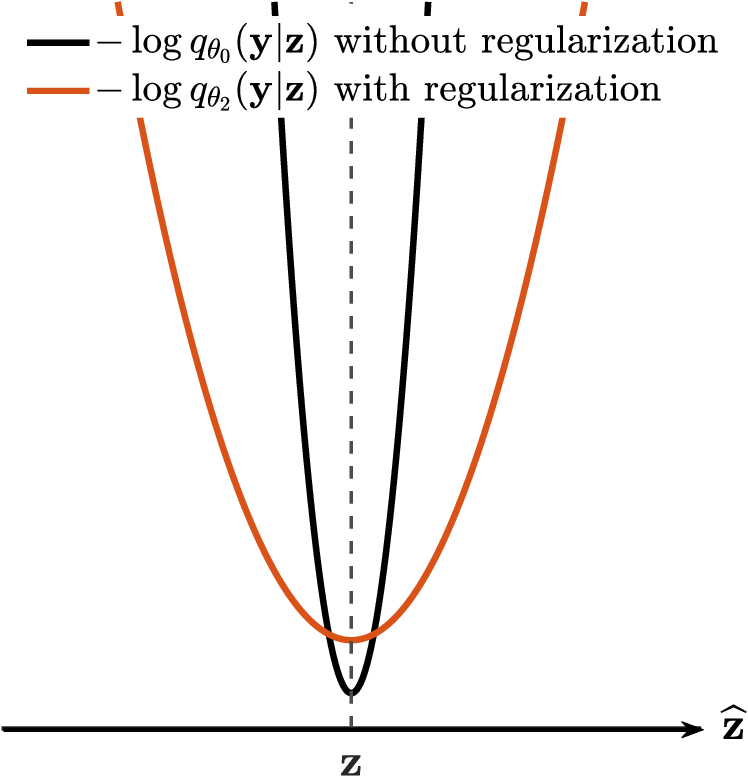}\label{fig:curvature_bad}}
        \caption{Comparison of the log posterior curvature smoothness under different channel conditions. (a) illustrates the smoothing effect of the proposed regularization under the good channel condition, while (b) shows the smoothing under the poor channel condition. $\theta_0$ denotes the parameter optimized without regularization, while $\theta_1$ and $\theta_2$ denote parameters optimized with regularization for small and large $\sigma^2$, respectively.}
        \label{fig:curvature}
    \end{figure}
    
    Finally, the regularization $\mathcal{R}(\boldsymbol{\phi},\boldsymbol{\theta})$ is obtained by taking the expectation with respect to the input data distribution $p(\mathbf{x})$, as follows:
    \begin{align}\label{eq:regularizer}
        \mathcal{R}(\boldsymbol{\phi},\boldsymbol{\theta})&=\mathbb{E}_{p_{\boldsymbol{\phi}}(\mathbf z)}\left[\mathcal{R}(\boldsymbol{\phi},\boldsymbol{\theta},\mathbf{z})\right]\\
        &=\mathbb{E}_{p(\mathbf x)}\left[\frac{\sigma^2}{2}\operatorname{Tr}\left(\mathcal{I}(f_{\boldsymbol{\phi}}(\mathbf{x}))\right)\right]\\
        &\simeq\frac{1}{N}\sum_{i=1}^N \frac{\sigma^2}{2}\operatorname{Tr}\left(\mathcal{I}(f_{\boldsymbol{\phi}}(\mathbf{x}^{(i)}))\right),\label{eq:reg_practice}
    \end{align}
    where~\eqref{eq:reg_practice} follows from Monte Carlo sampling, and $\{\mathbf{x}^{(i)}\}_{i=1}^N$ denotes samples drawn from the input distribution $p(\mathbf{x})$.

    Given the original loss function $\mathcal{L}(\boldsymbol{\phi},\boldsymbol{\theta})$, we can enhance robustness by employing the following proposed loss function:
    \begin{equation}
        \mathcal{L}(\boldsymbol{\phi},\boldsymbol{\theta}) + \lambda\mathcal{R}(\boldsymbol{\phi},\boldsymbol{\theta}),
    \end{equation}
    where the hyperparameter $\lambda$ controls the trade-off between the original loss and the proposed regularization term. 
    For instance, if the original loss is the cross-entropy loss, the modified loss becomes
    \begin{equation}
        \mathbb{E}_{p(\mathbf{x},\mathbf{y})}\Big[\mathbb{E}_{p_{\boldsymbol{\phi}}(\widehat{\mathbf{z}}|\mathbf{x})} \Big[-\log q_{\boldsymbol{\theta}}(\mathbf{y}|\widehat{\mathbf{z}})\Big]+\lambda 
        \frac{\sigma^2}{2}\operatorname{Tr}(\mathcal{I}(f_{\boldsymbol{\phi}}(\mathbf{x})))\Big].
    \end{equation}
    In practice, given a dataset $\{(\mathbf{x}^{(i)},\mathbf{y}^{(i)})\}_{i=1}^N$, this expectation can be computed using Monte Carlo sampling as follows:
    \begin{align}
    \frac{1}{N}\sum_{i=1}^N &  \left\{  - \frac{1}{L}\sum_{l=1}^L \log q_{\boldsymbol{\theta}}(\mathbf{y}^{(i)}|\widehat{\mathbf{z}}^{(i,l)}) \right. \nonumber \\
    & 
    \left.+\lambda \frac{\sigma^2}{2}\operatorname{Tr}(\mathcal{I}(f_{\boldsymbol{\phi}}(\mathbf{x}^{(i)})))\right\},
    \end{align}
    where $\widehat{\mathbf{z}}^{(i,l)}=f_{\boldsymbol{\phi}}(\mathbf{x}^{(i)})+\mathbf{n}^{(i,l)}$ with $\mathbf{n}^{(i,l)}\sim \mathcal{N}(0,\sigma^2I)$.  
    Note that the cross-entropy loss term is averaged over the channel noise, whereas the regularization term inherently incorporates it (see Proposition~\ref{prop:expected_KL}).

    \begin{figure}[t] 
        \centering
        \subfloat[][Conventional method]{
        \centering
        \begin{tabular}{c}
             \includegraphics[width=0.45\linewidth]{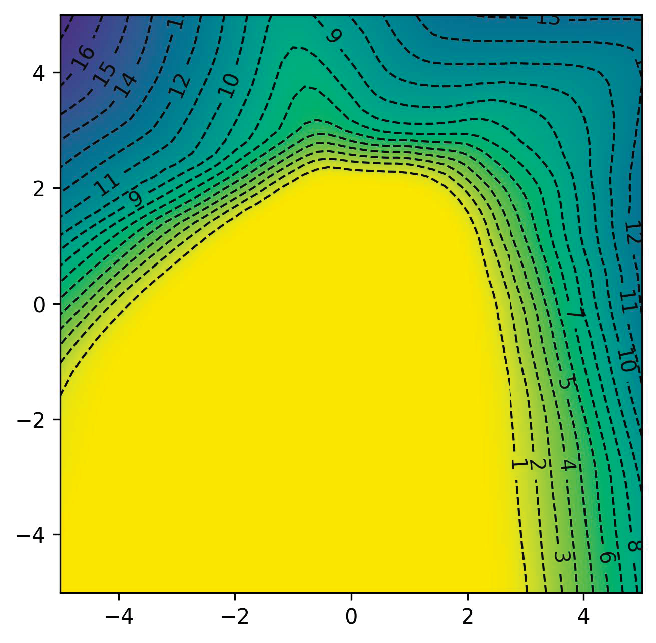}\\
             \includegraphics[width=0.45\linewidth]{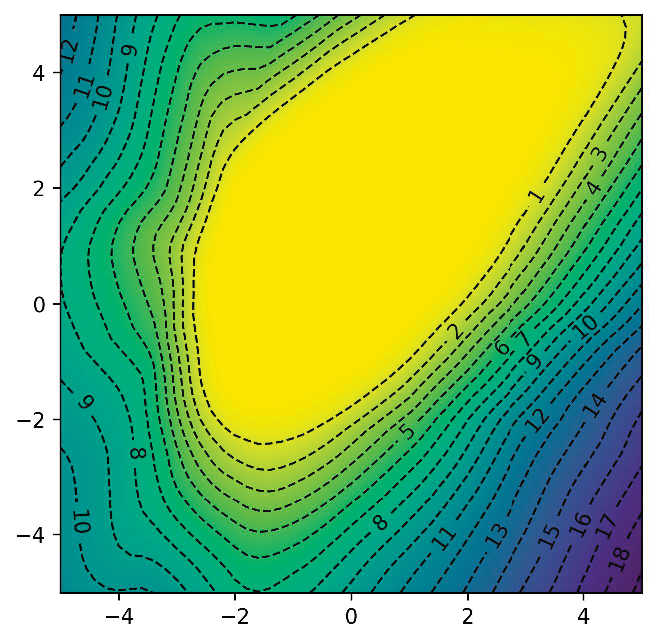}\\
             \includegraphics[width=0.45\linewidth]{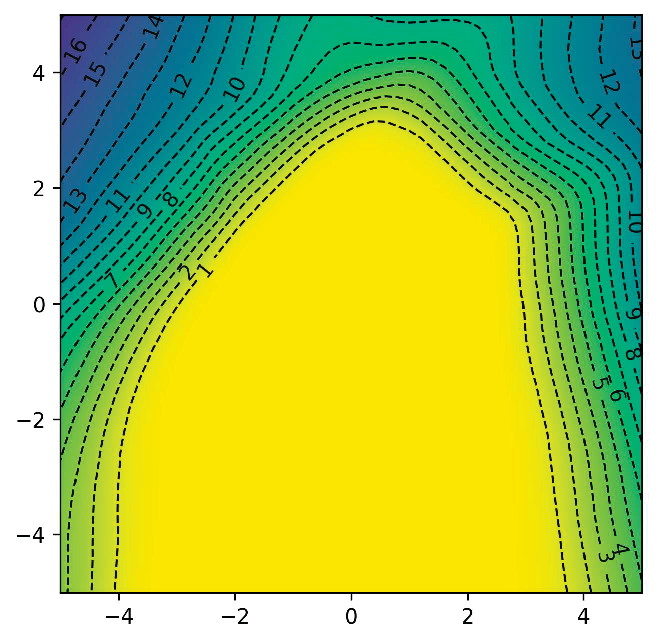}
        \end{tabular}
        \label{fig:posterior_deepjscc}}
        \subfloat[][Proposed method]{
        \centering
        \begin{tabular}{c}
             \includegraphics[width=0.45\linewidth]{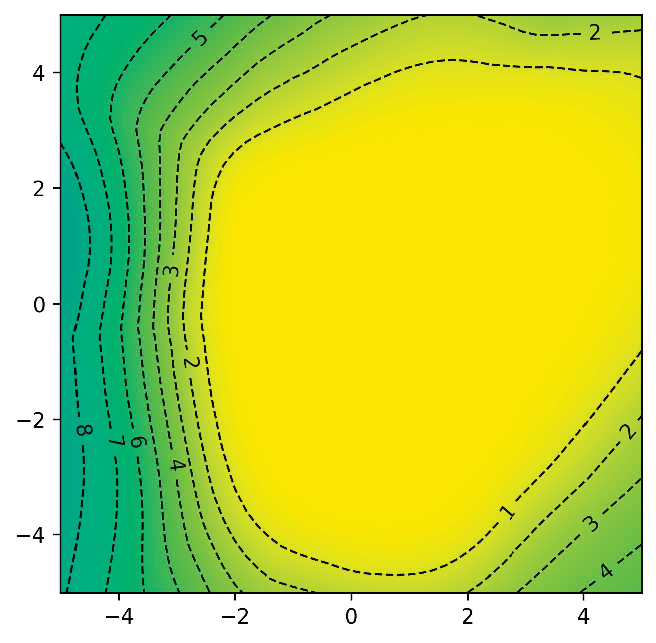}\\
             \includegraphics[width=0.45\linewidth]{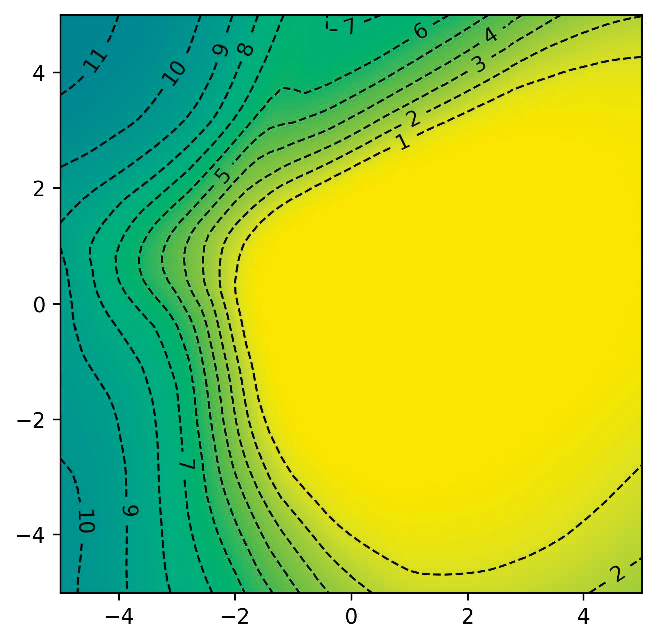}\\
             \includegraphics[width=0.45\linewidth]{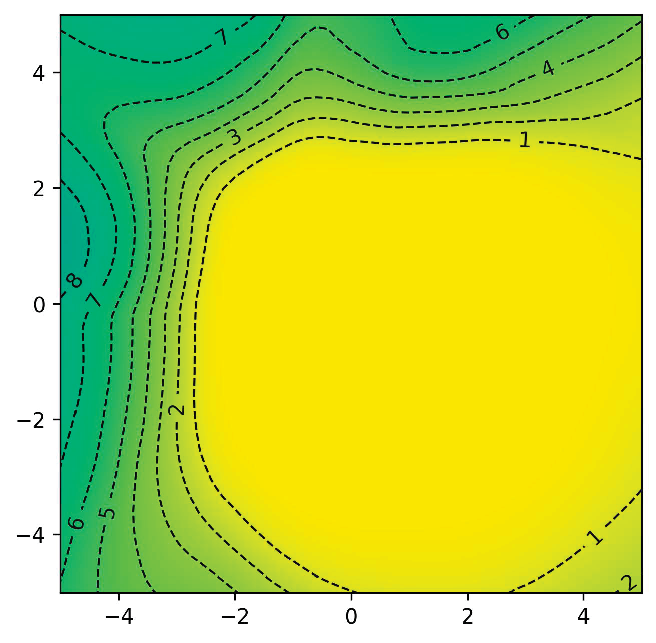}
        \end{tabular}
        \label{fig:posterior_proposed}}     
        \caption{Visualization of negative log posterior for a true label, comparing the conventional method and the proposed method. 
        DeepJSCC~\cite{Bourtsoulatze2019deep} is chosen to be a baseline JSCC model for a classification task, implemented with the architecture from~\cite{Shao2022learning,Shao2022vl-vfe} and trained on the CIFAR-10 dataset~\cite{Krizhevsky2009learning}.
        Each row corresponds to results obatined from the same input image. (a) shows the negative log posterior for DeepJSCC, and (b) shows the negative log posterior for the proposed model.}
        \label{fig:vis_posterior}
        \vspace{-2mm}
    \end{figure} 
    
    \subsection{Proposed Regularizer and Log-Posterior Curvature}

    We investigate the effect of regularization on the posterior distribution.
    Minimizing the regularization term smooths the curvature of the log posterior, as the Fisher information matrix in~\eqref{eq:cond_reg} is equivalent to the expected negative Hessian of $\log q_{\boldsymbol{\theta}}(\mathbf{y}|\mathbf{z})$~\cite{Vantrees2004detection}.
    Consequently, the impact of the channel noise can be mitigated by the proposed regularizer. 
    As shown in Fig.~\ref{fig:curvature}, the proposed regularizer may slightly reduce the posterior distribution compared to the conventional method under noise-free conditions, i.e., when $\mathbf{\widehat{z}} = \mathbf{z}$. 
    However, it effectively mitigates the impact of channel noise on the posterior distribution, enhancing overall robustness. 
    
    Importantly, because the regularization term includes the channel noise variance $\sigma^2$, the degree of smoothing adapts to channel conditions, which is a key advantage of the proposed method. 
    Fig.~\ref{fig:curvature} shows how channel conditions affect the curvature of the log posterior.
    When the channel condition is favorable (i.e., smaller noise variance), as shown in Fig.~\ref{fig:curvature}\subref{fig:curvature_good}, the regularization term $\mathcal{R}(\boldsymbol{\phi},\boldsymbol{\theta},\mathbf{z})$ remains low without significant reduction in $\operatorname{Tr}(\mathcal{I}(\mathbf{z}))$.
    For instance, if the noise variance is zero, then the regularizer vanishes. 
    However, as shown in Fig.~\ref{fig:curvature}\subref{fig:curvature_bad}, when the channel condition deteriorates (i.e., larger noise variance), $\operatorname{Tr}(\mathcal{I}(\mathbf{z}))$ should be substantially reduced to maintain the same regularization value. 
    This reduction corresponds to stronger smoothing of the log posterior $\log q_{\boldsymbol{\theta}}(\mathbf{y}|\mathbf{z})$.     
    This relationship aligns with the intuition that greater smoothing is necessary to preserve task performance when the encoded representation is heavily perturbed by noise.

    Fig.~\ref{fig:vis_posterior} shows the effect of regularization on the actual posterior distribution.
    Heatmaps visualize the negative log posteriors for the true label in two dimensions, using the two most significant axes obtained via principal component analysis (PCA) of the encoded representation $\mathbf{z}$.
    The results show that the proposed regularization method effectively smooths the curvature of the log posterior.
    As the input moves farther from the origin--indicating increased channel noise in the encoded representation--the posterior in the proposed method with regularization changes more gradually compared to the conventional model without regularization.
    Since the posterior remains stable despite channel noise perturbations, the task performance that depends on the posterior becomes more robust. 
    
    \section{Experimental Results}\label{sec:results}
    
    In this section, we present the experimental results for the proposed regularization. 
    To evaluate its effectiveness, we implement it into recent learning-based JSCC schemes for classification tasks and assess its impact. 

    \subsection{Experiment Settings}
   
    In simulations, we impose constraints on the encoder's power and latency to reflect a practical communication environment.
    Specifically, for fair comparison with previous studies, we constrain the maximum power of the encoded representation $\mathbf{z}$ to $P$, i.e., $\max_i |z_i|^2\leq P$ following~\cite{Shao2022learning, Xie2023robust}.
    Consequently, the channel condition is determined by the peak signal-to-noise ratio (PSNR), defined as
    \begin{equation}
        \text{PSNR}=10\log\frac{P}{\sigma^2}.
    \end{equation}
    Assuming a fixed transmission rate, latency depends on the number of symbols transmitted per data unit. 
    Thus, it is directly determined by the dimension of the encoded representation $\mathbf{z}$.

    We use the CIFAR-10 and CIFAR-100 datasets~\cite{Krizhevsky2009learning} for the classification task.
    The CIFAR-10 dataset contains 10 classes, while the CIFAR-100 dataset has 100 classes.
    To evaluate the compatibility of the proposed regularization across diverse environments, we select multiple recent deep learning-based JSCC schemes as baselines.
    For CIFAR-10, we adopt the neural network architecture proposed in~\cite{Shao2022learning,Shao2022vl-vfe,Xie2023robust}.
    For CIFAR-100, we use the same model as for CIFAR-10, modifying only the output dimension of the last layer to accommodate 100 classes for direct comparison with previous studies. 
    The hyperparameters of learning-based JSCC models are optimized for high classification accuracy before applying the proposed method.
    
    \begin{figure*}[!t] 
        \centering
        \subfloat[][Train PSNR \SI{10}{\decibel}]{\includegraphics[width=0.3\textwidth]{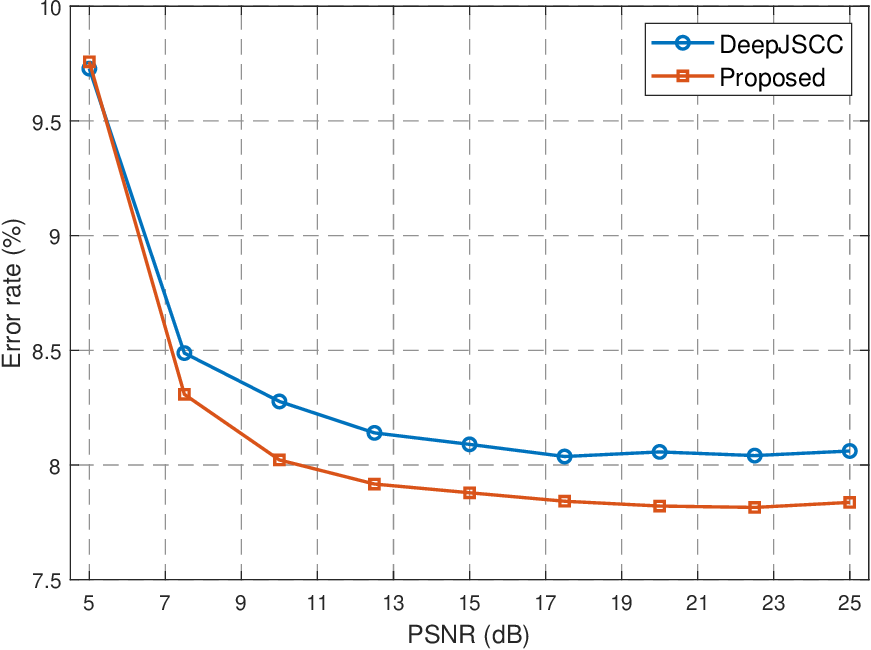}\label{fig:deepJSCC-10}}
        \hfil
        \subfloat[][Train PSNR \SI{15}{\decibel}]{\includegraphics[width=0.3\textwidth]{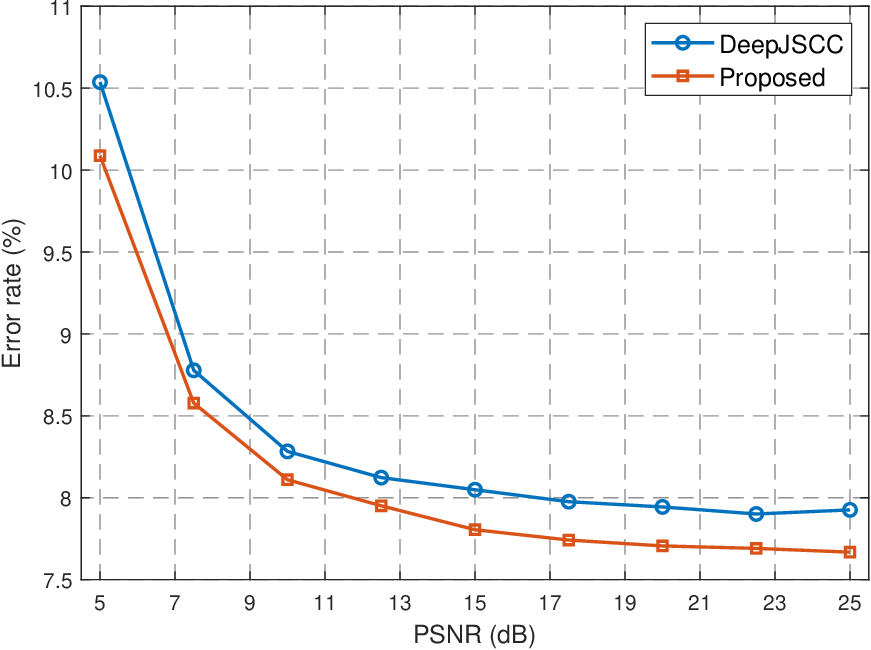}\label{fig:deepJSCC-15}}
        \hfil
        \subfloat[][Train PSNR \SI{20}{\decibel}]{\includegraphics[width=0.3\textwidth]{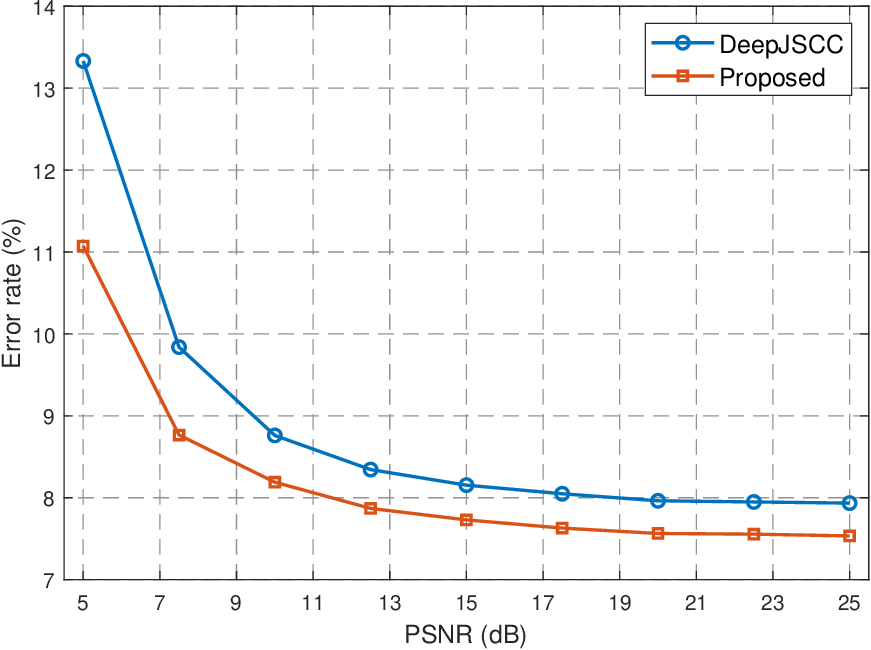}\label{fig:deepJSCC-20}}        
        \caption{Comparison of CIFAR-10 classification error rates between DeepJSCC and the proposed method across PSNRs ranging from \SI{5}{\decibel} to \SI{25}{\decibel}. Models in (a), (b), and (c) are trained with fixed channel PSNRs of \SI{10}{\decibel}, \SI{15}{\decibel}, and \SI{20}{\decibel}, respectively.}
        \label{fig:deepJSCC}
        \vspace{-4mm}
    \end{figure*}
    
    \begin{figure*}[!t] 
        \centering
        \subfloat[][Train PSNR \SI{10}{\decibel}]{\includegraphics[width=0.3\textwidth]{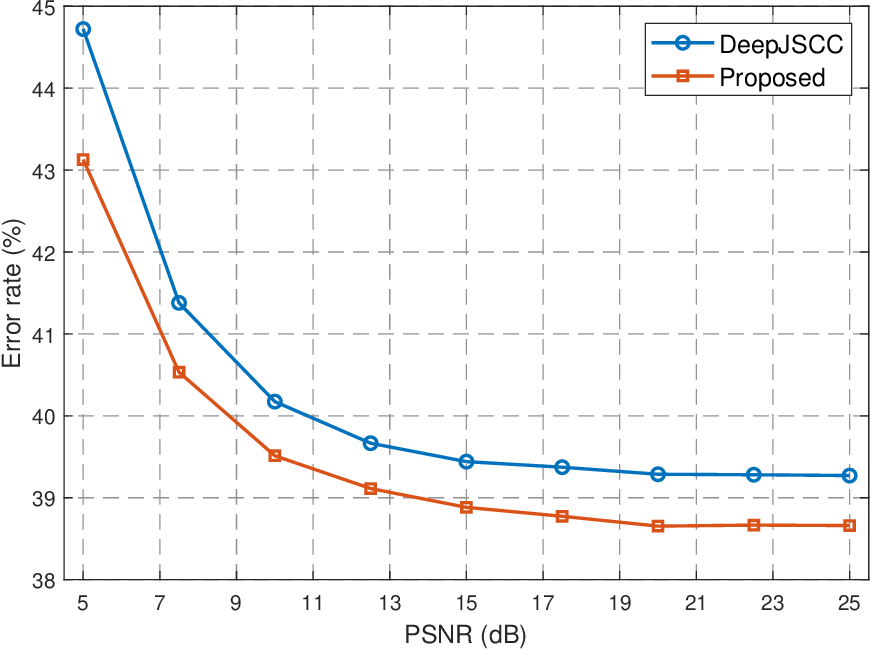}\label{fig:deepJSCC_c100-10}}
        \hfil
        \subfloat[][Train PSNR \SI{15}{\decibel}]{\includegraphics[width=0.3\textwidth]{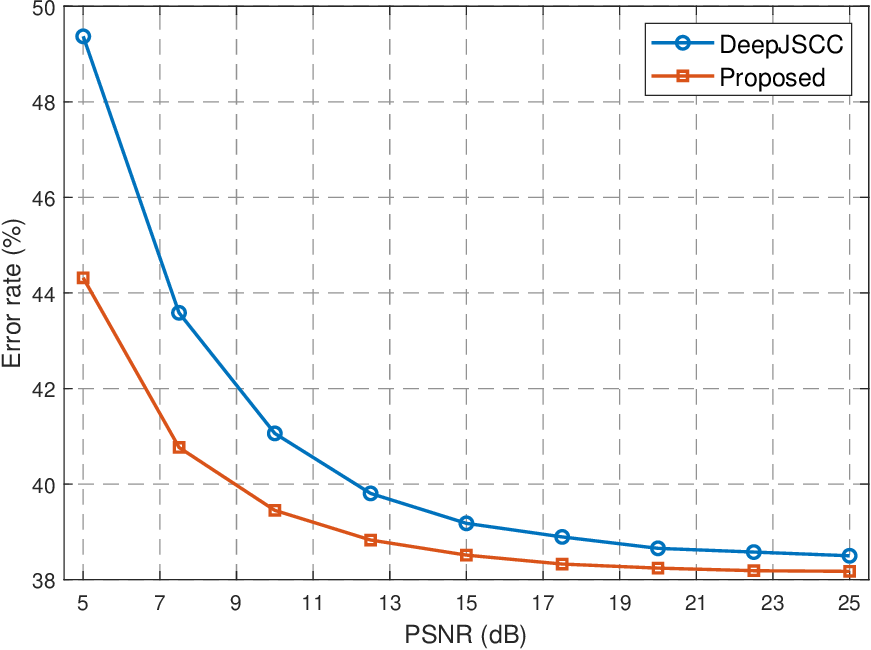}\label{fig:deepJSCC_c100-15}}
        \hfil
        \subfloat[][Train PSNR \SI{20}{\decibel}]{\includegraphics[width=0.3\textwidth]{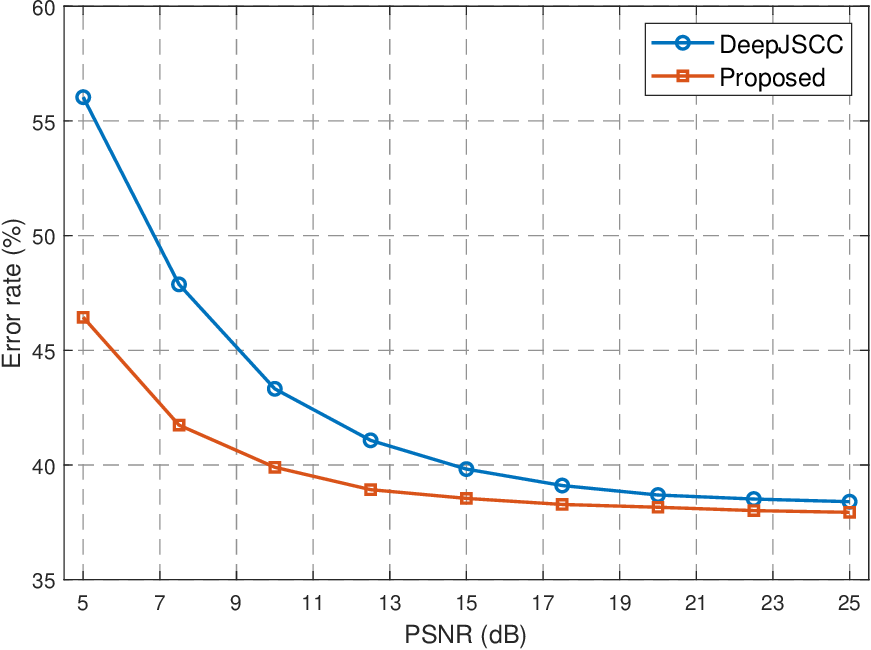}\label{fig:deepJSCC_c100-20}}        
        \caption{Comparison of CIFAR-100 classification error rates between DeepJSCC and the proposed method across PSNRs ranging from \SI{5}{\decibel} to \SI{25}{\decibel}. Models in (a), (b), and (c) are trained with fixed channel PSNRs of \SI{10}{\decibel}, \SI{15}{\decibel}, and \SI{20}{\decibel}, respectively.}
        \label{fig:deepJSCC_c100}
        \vspace{-4mm}
    \end{figure*}   
    
    \begin{itemize}
        \item \emph{DeepJSCC}~\cite{Bourtsoulatze2019deep}: DeepJSCC was originally proposed as a learning-based JSCC model for image recovery using an autoencoder architecture.
        To adapt it for image classification, the output layer is modified to match the number of classes and trained with the cross-entropy loss function, as in~\cite{Shao2022learning}. 
        The encoder and decoder architectures follow the structure described in~\cite{Shao2022learning, Shao2022vl-vfe}, with the encoded representation dimension $k=8$ for CIFAR-10 and $k=16$ for CIFAR-100.
        \item \emph{VL-VFE}~\cite{Shao2022learning}: VL-VFE is a learning-based JSCC model for image classification, based on the variational information bottleneck (VIB) principle. 
        It employs a log-uniform prior on the encoded representation to induce sparsity~\cite{Kingma2015variational}, pruning unnecessary dimensions for communication efficiency. 
        Additionally, VL-VFE adapts to dynamic channel conditions by adjusting the dimension of encoded representation based on the channel condition.
        The encoder and decoder architectures follow the structure in~\cite{Shao2022learning, Shao2022vl-vfe}, with the initial encoded representation dimension set to $64$ for both datasets.
        \item \emph{DT-JSCC}~\cite{Xie2023robust}: Discrete task-oriented JSCC (DT-JSCC) is a task-oriented communication scheme with digital modulation, built upon the robust information bottleneck (RIB) principle. 
        It introduces redundant information to enhance robustness against channel variations, balancing task-relevant and redundant information through the RIB principle. 
        Additionally, vector quantization and $M$-PSK digital modulation are employed to enable digital communication. 
        The encoder and decoder architectures follow the structure in~\cite{Xie2023robust}. 
        For vector quantization, the codebook size is set to $8$ for CIFAR-10 and $16$ for CIFAR-100, while the encoded dimension is fixed at $k=16$ for both datasets.
    \end{itemize}
    
    The proposed methods are implemented by incorporating the proposed regularization term into the loss functions of three representative learning-based JSCC schemes and training them accordingly.
    For models trained under a fixed channel PSNR, the regularization term in~\eqref{eq:cond_reg} can be simplified by omitting the variance, resulting in $\lambda\operatorname{Tr}(\mathcal{I}(\mathbf{z}))$, where the hyperparameter $\lambda\in[0.1, 1]$ is optimized via grid search.
    When the channel PSNR values are randomly sampled during training, the regularization term~\eqref{eq:cond_reg} is implemented as $\frac{\lambda\sigma^2}{2}\operatorname{Tr}(\mathcal{I}(\mathbf{z}))$, and the hyperparameter $\frac{\lambda\sigma^2}{2} \in [0.5, 1.5]$ is optimized using a grid search while setting $\sigma^2$ to \SI{10}{\decibel}.
    For DeepJSCC and VL-VFE, the regularization terms are computed for AWGN channels, whereas for DT-JSCC, the regularization term is computed using the extended channel model from~\cite{Xie2023robust}, which integrates digital modulation, demodulation, and the physical channel. 
    Since the proposed method introduces an additional term in the loss function, it increases computational complexity during training but does not affect inference complexity. 
    All models are trained for $320$ epochs using the Adam optimizer~\cite{Kingma2015adam}.
    
    \subsection{Performance Evaluation in Analog Semantic Communications}

    \begin{figure*}[t] 
        \centering
        \subfloat[][CIFAR-10]{\includegraphics[width=0.4\textwidth]{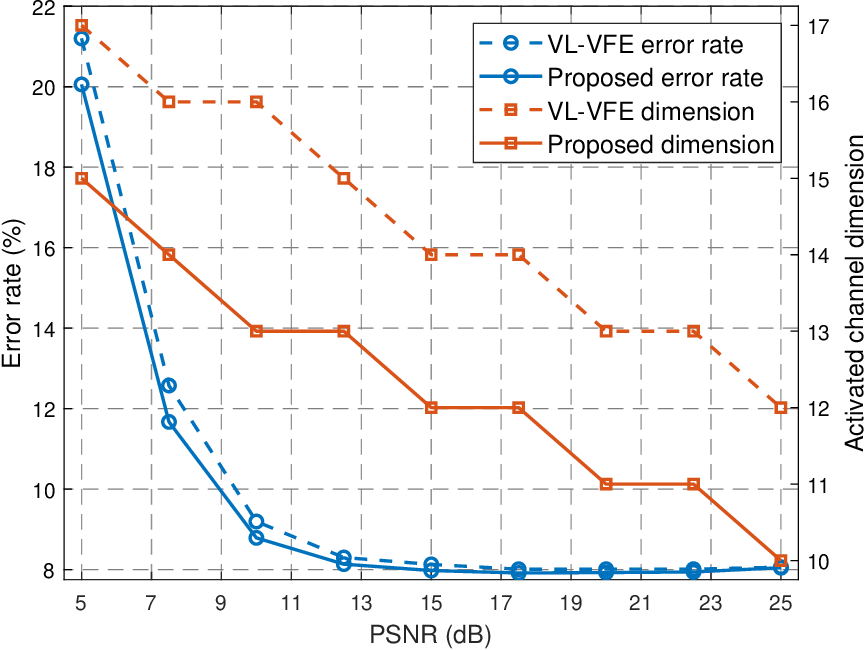}\label{fig:vl_vfe_c10}}
        \hfil
        \subfloat[][CIFAR-100]{\includegraphics[width=0.4\textwidth]{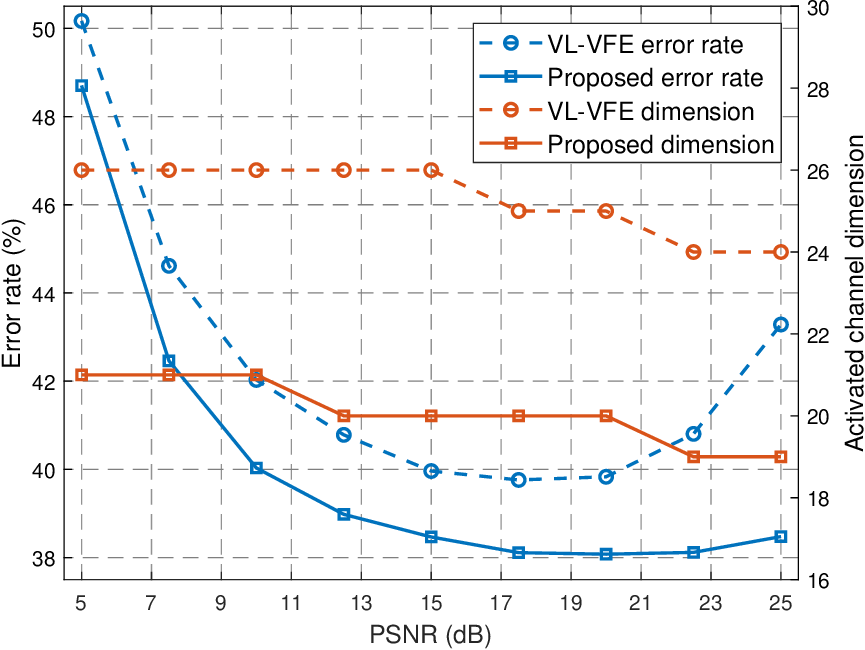}\label{fig:vl_vfe_c100}}        
        \caption{Comparison of classification error rates and activated channel dimensions between VL-VFE~\cite{Shao2022learning} and the proposed method on the CIFAR-10 and CIFAR-100 datasets. Models in (a), (b) are trained with varying channel PSNRs from \SI{10}{\decibel} to \SI{25}{\decibel}.}
        \label{fig:vl_vfe}
        \vspace{-4mm}
    \end{figure*}
    
    In this subsection, we evaluate the performance of learning-based JSCC schemes with analog communications, specifically DeepJSCC~\cite{Bourtsoulatze2019deep} and VL-VFE~\cite{Shao2022learning}.
    We compare the classification error rate of DeepJSCC~\cite{Bourtsoulatze2019deep} and the proposed method on the CIFAR-10 and CIFAR-100 datasets, as shown in Fig.~\ref{fig:deepJSCC} and Fig.~\ref{fig:deepJSCC_c100}.
    Both DeepJSCC and the proposed models are trained with fixed channel PSNR values of \SI{10}{\decibel}, \SI{15}{\decibel}, and \SI{20}{\decibel}, as in~\cite{Bourtsoulatze2019deep}.
    To assess robustness when the training and testing PSNRs differ, we evaluate the models over a broader PSNR range from \SI{5}{\decibel} to \SI{25}{\decibel}.
    The results show that the proposed regularization consistently improves classification accuracy across all PSNR values for both datasets. 
    Notably, the proposed method significantly outperforms DeepJSCC when trained at high SNR and tested in low SNR, as shown in Fig.~\ref{fig:deepJSCC}\subref{fig:deepJSCC-20} and Fig.~\ref{fig:deepJSCC_c100}\subref{fig:deepJSCC_c100-20}.
    This indicates that the proposed regularization can effectively enhance robustness in challenging channel conditions unseen during training.

\begin{figure*}[!t] 
        \centering
        \subfloat[][Train PSNR \SI{15}{\decibel}]{\includegraphics[width=0.3\textwidth]{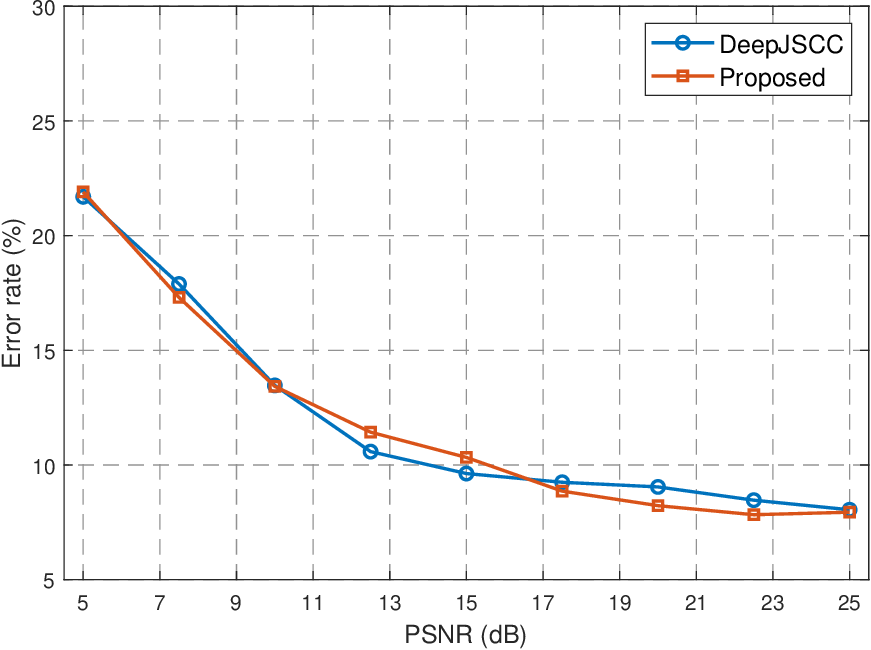}\label{fig:deepJSCC-fading-15}}
        \hfil
        \subfloat[][Train PSNR \SI{20}{\decibel}]{\includegraphics[width=0.3\textwidth]{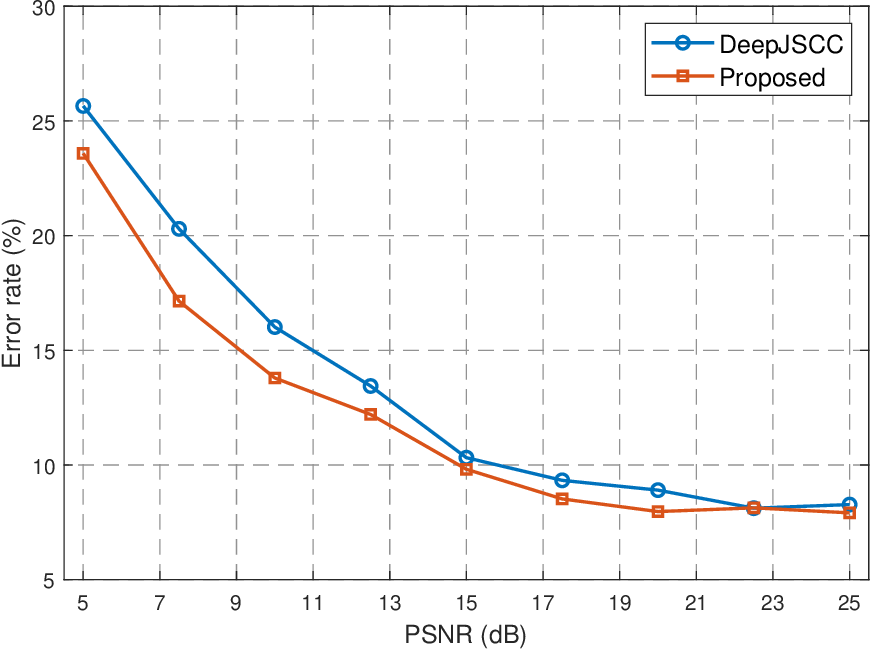}\label{fig:deepJSCC-fading-20}}
        \hfil
        \subfloat[][Train PSNR \SI{25}{\decibel}]{\includegraphics[width=0.3\textwidth]{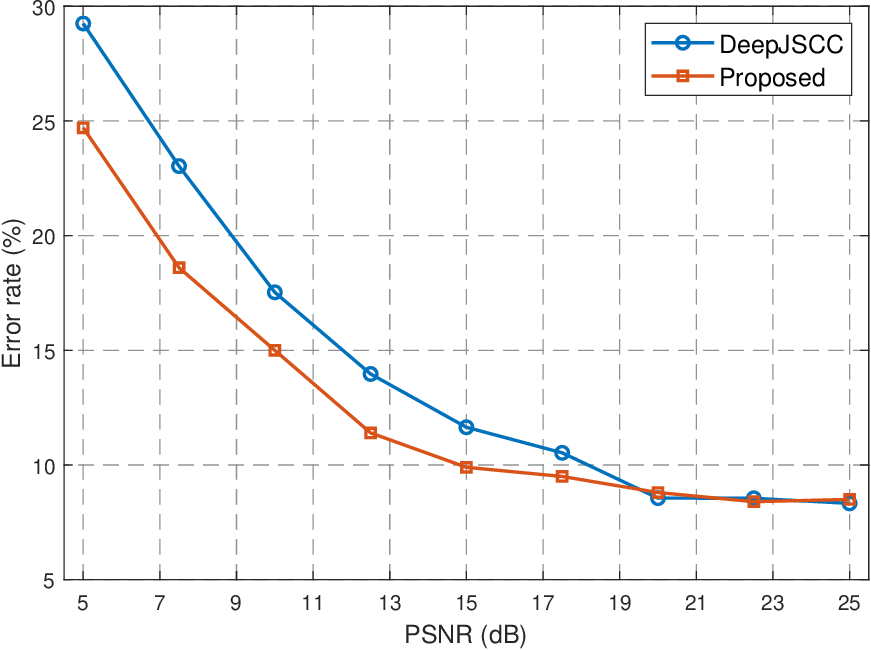}\label{fig:deepJSCC-fading-25}}      
        \caption{Comparison of CIFAR-10 classification error rates between DeepJSCC and the proposed method in a Rayleigh fading channel across PSNRs ranging from \SI{5}{\decibel} to \SI{25}{\decibel}. Models in (a), (b), and (c) are trained on an AWGN channel with fixed channel PSNR values of \SI{15}{\decibel}, \SI{20}{\decibel}, and \SI{25}{\decibel}, respectively.}
        \label{fig:deepJSCC-fading}
        \vspace{-4mm}
    \end{figure*}
    
    \begin{figure*}[!t] 
        \centering
        \subfloat[][Train PSNR \SI{15}{\decibel}]{\includegraphics[width=0.3\textwidth]{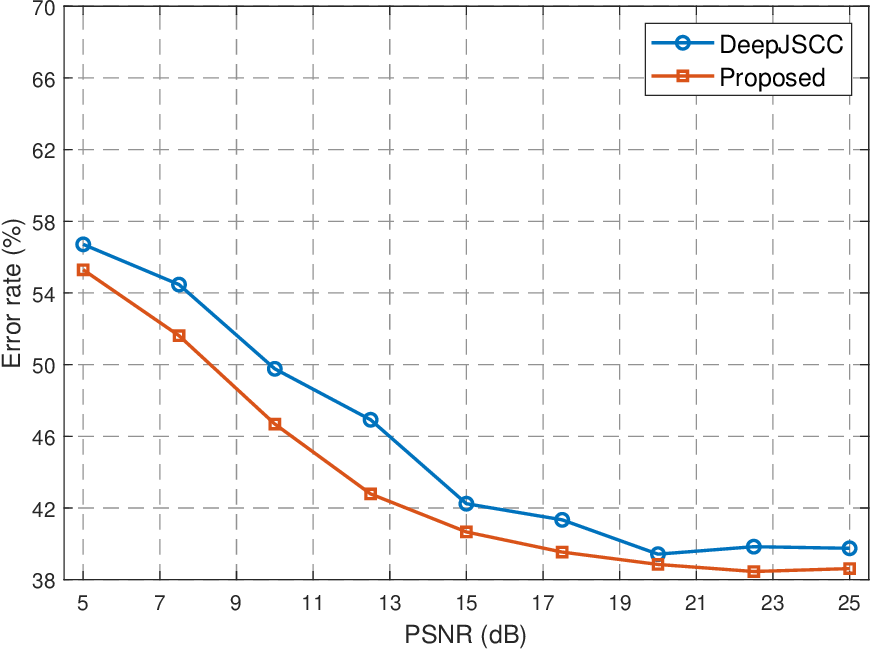}\label{fig:deepJSCC-fading_c100-10}}
        \hfil
        \subfloat[][Train PSNR \SI{20}{\decibel}]{\includegraphics[width=0.3\textwidth]{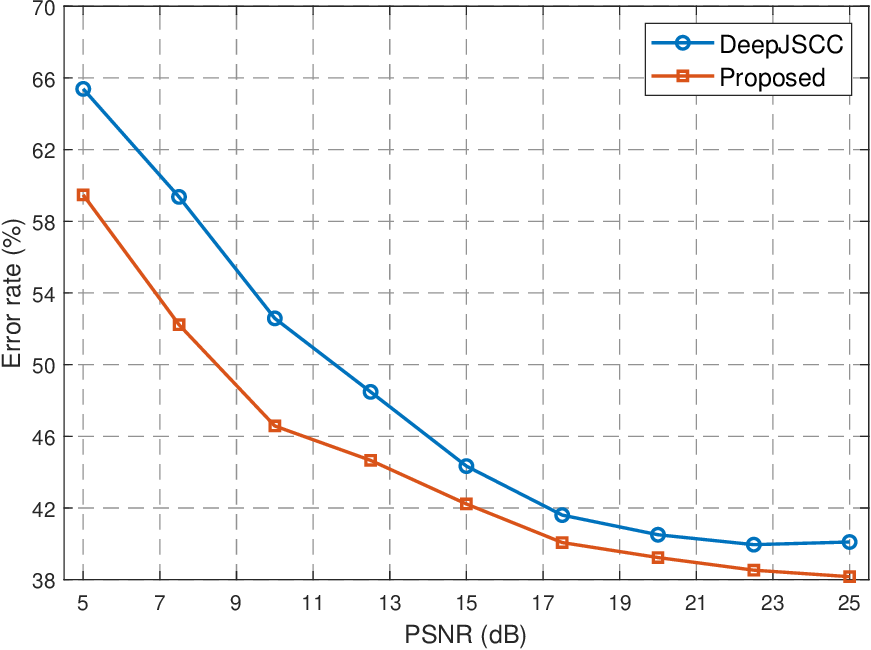}\label{fig:deepJSCC-fading_c100-15}}
        \hfil
        \subfloat[][Train PSNR \SI{25}{\decibel}]{\includegraphics[width=0.3\textwidth]{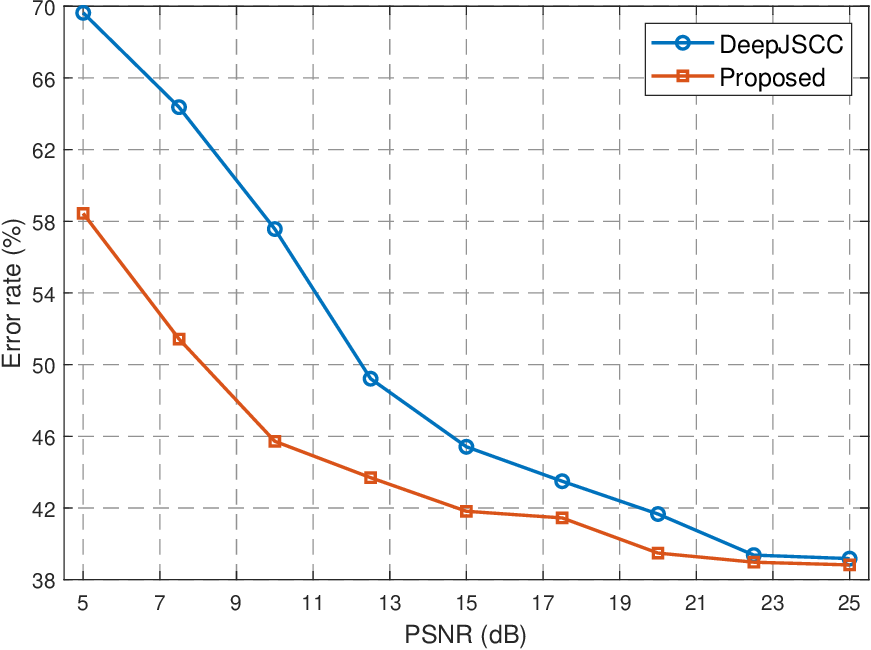}\label{fig:deepJSCC-fading_c100-20}}        
        \caption{Comparison of CIFAR-100 classification error rates between DeepJSCC and the proposed method in a Rayleigh fading channel across PSNRs ranging from \SI{5}{\decibel} to \SI{25}{\decibel}. Models in (a), (b), and (c) are trained on an AWGN channel with fixed PSNR values of \SI{15}{\decibel}, \SI{20}{\decibel}, and \SI{25}{\decibel}, respectively.}
        \label{fig:deepJSCC-fading_c100}
    \end{figure*}    

    Next, we compare the classification error rate and the activated channel dimension of VL-VFE~\cite{Shao2022learning} and the proposed method on the CIFAR-10 and CIFAR-100 datasets, as shown in Fig.~\ref{fig:vl_vfe}.
    Both VL-VFE and the proposed models are trained under varying channel conditions, with the training PSNR uniformly sampled from \SI{10}{\decibel} to \SI{25}{\decibel} for each batch, following~\cite{Shao2022learning}.
    For a fair comparison, the hyperparameter $\lambda$ in the proposed method is chosen to ensure that the activated channel dimension remains lower than that of VL-VFE across the entire range of testing PSNRs.
    The evaluation results show that the proposed method consistently achieves higher classification accuracy while maintaining lower activated channel dimensions for both datasets.
    Notably, for CIFAR-100, the classification error rate of VL-VFE increases as the PSNR rises from \SI{20}{\decibel} to \SI{25}{\decibel}, whereas the proposed method maintains a relatively stable error rate. 
    These results indicate that the proposed regularization effectively ensures robustness when trained with varying noise variances.

    To evaluate the robustness of JSCC models under changes in the channel statistical model, we test DeepJSCC and the proposed method on a Rayleigh fading channel after training them on an AWGN channel, as shown in Fig.~\ref{fig:deepJSCC-fading} and Fig.~\ref{fig:deepJSCC-fading_c100}.
    We consider a slow fading channel where the channel coefficient $h$ remains constant during the transmission of a single encoded representation.
    Assuming perfect channel estimation and equalization, the received representation is given by $\widehat{\mathbf{z}}=\mathbf{z}+\mathbf{n}/|h|$ where $h \sim \mathcal{CN}(0,1)$, and the PSNR is computed as $10\log(P/\sigma^2)$.
    The results show that the proposed method consistently outperforms DeepJSCC across all PSNR values for both datasets. 
    Notably, the proposed method maintains relatively stable classification accuracy as the training PSNR varies, whereas the error rate of DeepJSCC without regularization increases significantly at higher training PSNRs.
    These finding indicates that the proposed regularization enhances robustness against channel variations, even when the statistical model of the testing channel differs from that of the training channel.
   
    \subsection{Performance Evaluation in Digital Semantic Communications}
    
\begin{figure*}[t] 
        \centering
        \subfloat[][Train PSNR \SI{10}{\decibel}]{\includegraphics[width=0.4\textwidth]{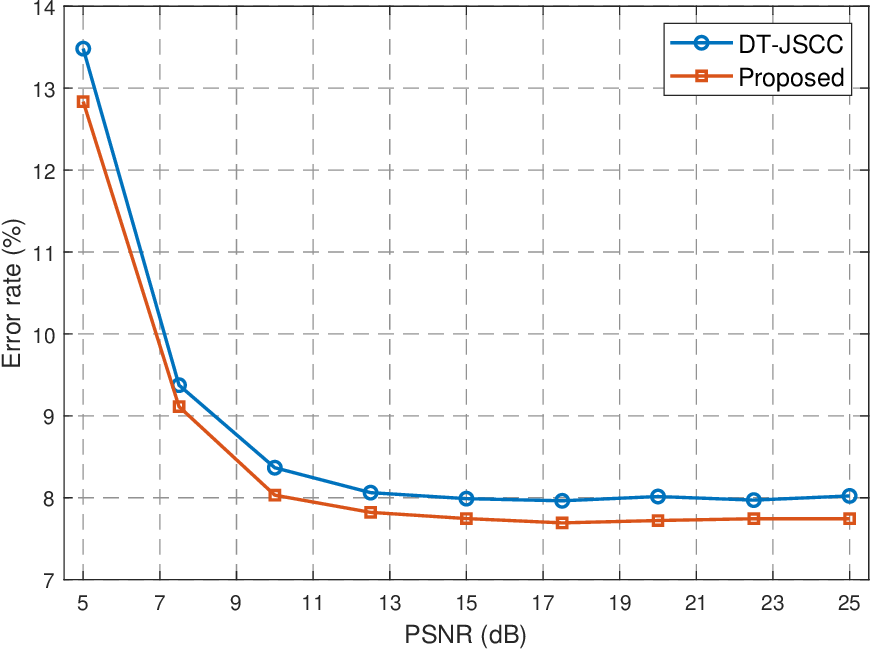}\label{fig:dt_jscc_10dB}}
        \hfil
        \subfloat[][Train PSNR \SI{20}{\decibel}]{\includegraphics[width=0.4\textwidth]{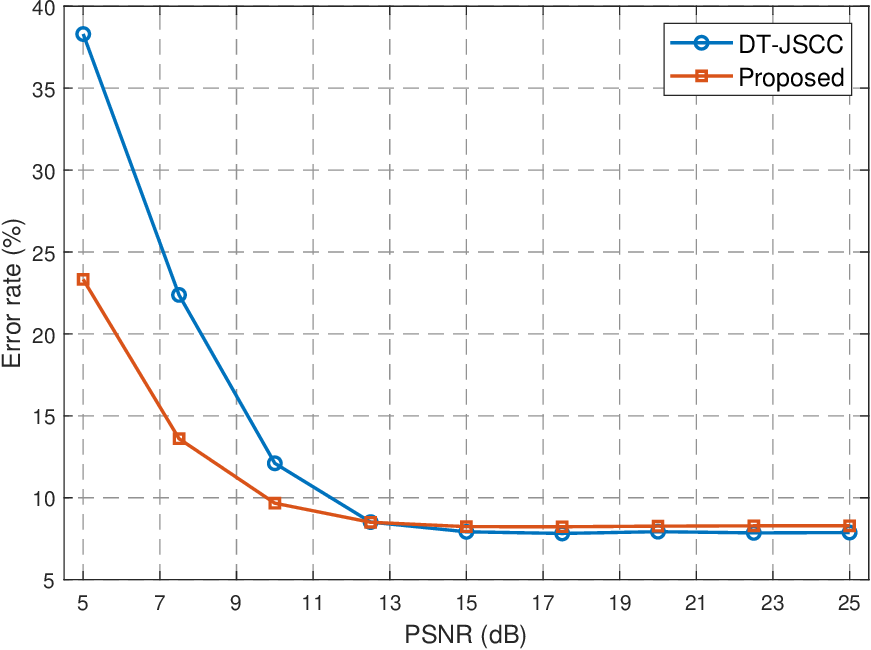}\label{fig:dt_jscc_20dB}}
        
        \caption{Comparison of CIFAR-10 classification error rate between DT-JSCC~\cite{Xie2023robust} and the proposed method across PSNRs ranging from \SI{5}{\decibel} to  \SI{25}{\decibel}. Models in (a) and (b) are trained with fixed channel PSNRs of \SI{10}{\decibel} and \SI{20}{\decibel}, respectively.}
        \label{fig:dt_jscc}
        \vspace{-4mm}
    \end{figure*}

\begin{figure*}[t] 
        \centering
        \subfloat[][Train PSNR \SI{15}{\decibel}]{\includegraphics[width=0.4\textwidth]{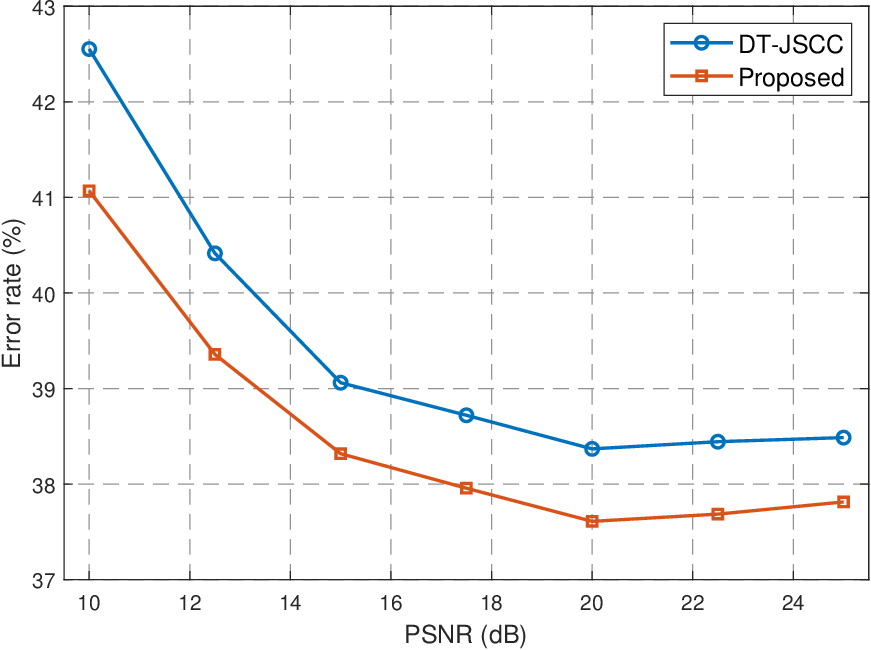}\label{fig:dt_jscc_c100_15dB}}
        \hfil
        \subfloat[][Train PSNR \SI{25}{\decibel}]{\includegraphics[width=0.4\textwidth]{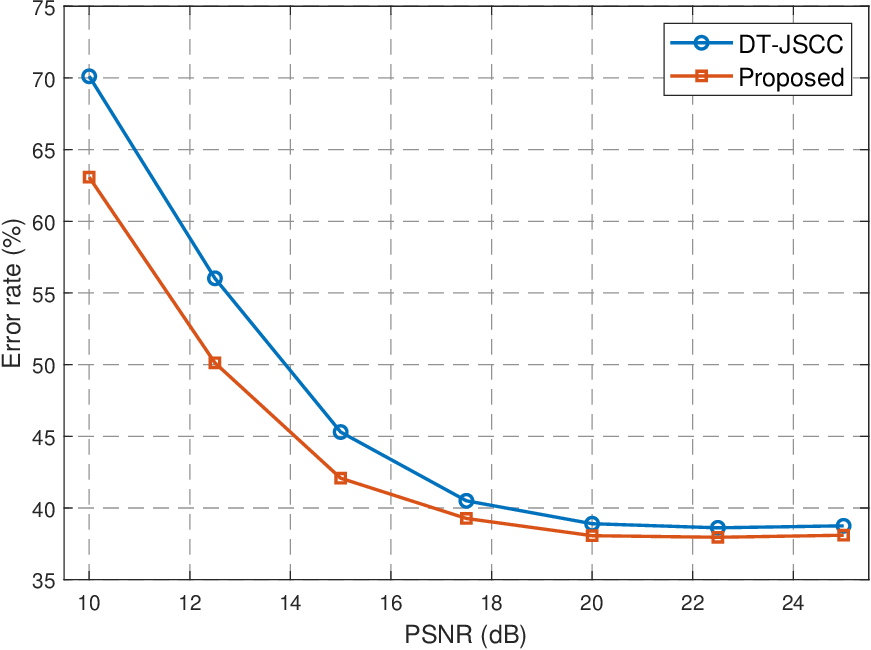}\label{fig:dt_jscc_c100_25dB}}        
        \caption{Comparison of CIFAR-100 classification error rate between DT-JSCC~\cite{Xie2023robust} and the proposed method across PSNRs ranging from \SI{10}{\decibel} to  \SI{25}{\decibel}. 
        Models in (a) and (b) are trained with fixed channel PSNRs of \SI{15}{\decibel} and \SI{25}{\decibel}, respectively.}
        \label{fig:dt_jscc_c100}
        \vspace{-4mm}
    \end{figure*}

    This subsection evaluates the effectiveness of the proposed regularization in digital semantic communications. 
    We compare the classification error rates of DT-JSCC~\cite{Xie2023robust} and the proposed method on the CIFAR-10 and CIFAR-100 datasets, as shown in Fig.~\ref{fig:dt_jscc} and Fig.~\ref{fig:dt_jscc_c100}.
    Both DT-JSCC and the corresponding proposed method are trained with fixed PSNR values of \SI{15}{\decibel} and \SI{25}{\decibel}, then tested across a range of PSNR values from \SI{10}{\decibel} to \SI{25}{\decibel}.
    The results show that the proposed regularization is effective even for discrete representations $\mathbf{z}$, enhancing classification accuracy compared to DT-JSCC without regularization. 
    Consistent with the findings from analog communication, the proposed regularization provides significant performance gains when models trained at high PSNRs are tested at lower PSNRs.

    \begin{figure}[t] 
        \centering
        \includegraphics[width=0.4\textwidth]{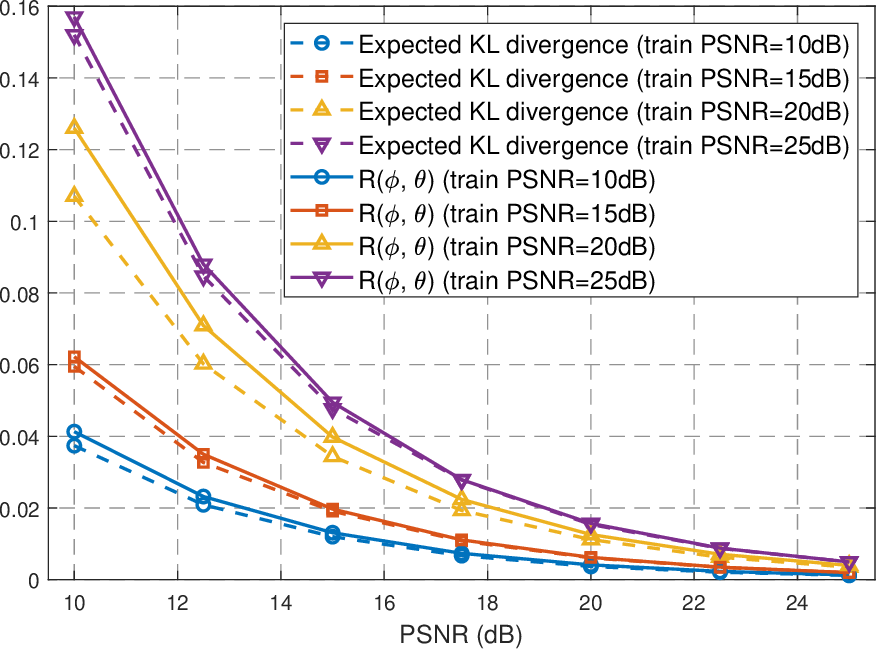}
        \caption{Comparison of the expected KL divergence and the proposed regularization term for the DeepJSCC on the CIFAR-10 dataset. Models are trained at \SI{10}{\decibel}, \SI{15}{\decibel}, \SI{20}{\decibel}, and \SI{25}{\decibel}. The expected KL divergence refers the term $\mathbb{E}\left[D_{KL}\left(q_{\boldsymbol{\theta}}(\mathbf{y}|\mathbf{z})\middle\|q_{\boldsymbol{\theta}}(\mathbf{y}|\widehat{\mathbf{z}})\right)\right]$, and it is computed by taking the empirical mean from sampling channel noise 20 times for each data samples.}
        \label{fig:kl_vanilla}
    \end{figure}

    \begin{figure}[t] 
        \centering
        \includegraphics[width=0.4\textwidth]{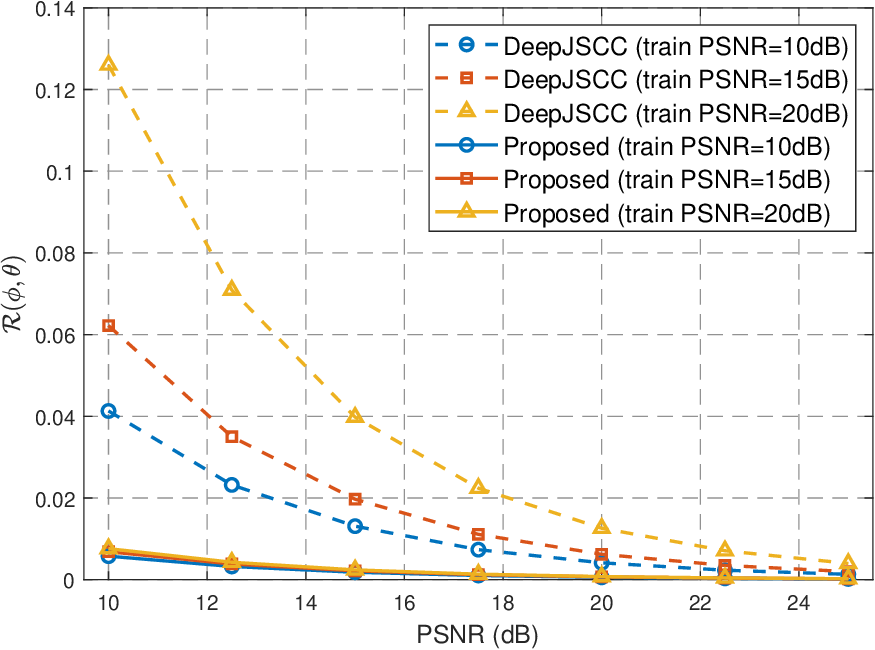}
        \caption{Comparison of $\mathcal{R}(\boldsymbol{\phi},\boldsymbol{\theta})$ values between DeepJSCC and the proposed method on the CIFAR-10 dataset. Models are trained at \SI{10}{\decibel}, \SI{15}{\decibel}, and \SI{20}{\decibel}.}
        \label{fig:kl_reg}
        \vspace{-4mm}
    \end{figure}
  
    \subsection{Comparative Analysis}
    
    This subsection presents a comparative analysis of robustness based on DeepJSCC.
    Fig.~\ref{fig:kl_vanilla} compares the expected KL divergence $\mathbb{E}\left[D_{KL}\left(q_{\boldsymbol{\theta}}(\mathbf{y}|\mathbf{z})\middle\|q_{\boldsymbol{\theta}}(\mathbf{y}|\widehat{\mathbf{z}})\right)\right]$ and the regularization term $\mathcal{R}(\boldsymbol{\phi},\boldsymbol{\theta})= \frac{\sigma^2}{2}\operatorname{Tr}\left(\mathcal{I}(\mathbf{z})\right)$ for DeepJSCC models. 
    Since the expected KL divergence cannot be computed in closed form, it is estimated using a sample average over noise vectors.
    The results show that the discrepancy between the KL divergence and the proposed regularization term remains small across the entire range of train PSNR and test PSNR values, numerically validating the accuracy of the Taylor approximation in Proposition~\ref{prop:expected_KL}.
    Moreover, the numerical results indicate that training at low PSNR values reduces both the KL divergence and the regularization term, thereby enhancing model robustness.
    
    Fig.~\ref{fig:kl_reg} compares the values of $\mathcal{R}(\boldsymbol{\phi},\boldsymbol{\theta}) = \frac{\sigma^2}{2}\operatorname{Tr}\left(\mathcal{I}(\mathbf{z})\right)$ for DeepJSCC trained without regularization and the proposed models trained with regularization.
    In DeepJSCC, this term varies substantially depending on the training PSNR and becomes smaller as the training PSNR decreases.
    This observation indicates that end-to-end training on noisier channels enhances robustness against channel noise. 
    This aligns with findings in~\cite{Xie2023robust}, which report that the learning-based JSCC models achieve lower error rates when trained at lower PSNRs.
    The proposed method, in contrast, maintains relatively stable values of $\mathcal{R}(\boldsymbol{\phi},\boldsymbol{\theta})$ across different training PSNR values, consistently lower than those of DeepJSCC over the entire testing PSNR range.
    This implies that the proposed regularization strategy effectively enhances robustness beyond conventional end-to-end training with a noisy channel. 
       
    \section{Conclusion}\label{sec:conclusion}
    
    In this paper, we proposed a novel regularization method to enhance robustness in learning-based JSCC schemes. 
    We formulated a novel regularization term based on the expected KL divergence between noise-free and noisy posteriors.
    Using a second order Taylor approximation, we showed that the expectation of the KL divergence conditioned on the encoded representation can be analytically approximated using the Fisher information matrix and the noise covariance matrix. 
    Furthermore, we revealed the regularization's smoothing impact on the posterior distribution, which adapts to the channel conditions.
    We integrated the proposed regularization into recent learning-based JSCC schemes and validated its effectiveness and compatibility through extensive experiments.
    Since the proposed regularization is architecture-agnostic, it can be widely applied to other semantic communication schemes that rely on posterior distributions for task execution. 

    \appendices
    \section{Proof of Proposition 1}\label{sec:proof1}
    Using the second order Taylor approximation, the KL divergence in~\eqref{eq:KLdiv} can be approximated as follows:
    \begin{align}\label{eq:KLdiv_Taylor_expansion}
        &D_{KL}\left(q_{\boldsymbol{\theta}}(\mathbf{y}|\mathbf{z})\middle\|q_{\boldsymbol{\theta}}(\mathbf{y}|\widehat{\mathbf{z}})\right) \nonumber \\
        &\approx D_{KL}\left(q_{\boldsymbol{\theta}}(\mathbf{y}|\mathbf{z})\middle\|q_{\boldsymbol{\theta}}(\mathbf{y}|\mathbf{z})\right)\nonumber \\
        &+ \left .\nabla_{\widehat{\mathbf{z}}}D_{KL}\left(q_{\boldsymbol{\theta}}(\mathbf{y}|\mathbf{z})\middle\|q_{\boldsymbol{\theta}}(\mathbf{y}|\widehat{\mathbf{z}})\right)\right|_{\widehat{\mathbf{z}}=\mathbf{z}}^{\mathsf{T}}(\widehat{\mathbf{z}}-\mathbf{z}) \nonumber \\
        &+\frac{1}{2}(\widehat{\mathbf{z}}-\mathbf{z})^\mathsf T\left .\nabla_{\widehat{\mathbf{z}}}^2 D_{KL}\left(q_{\boldsymbol{\theta}}(\mathbf{y}|\mathbf{z})\middle\|q_{\boldsymbol{\theta}}(\mathbf{y}|\widehat{\mathbf{z}})\right)\right |_{\widehat{\mathbf{z}}=\mathbf{z}}(\widehat{\mathbf{z}}-\mathbf{z}).
    \end{align}
    The zero order term is eliminated by the definition of KL divergence. 
    The first order term also becomes zero since
    \begin{align}
    &\left .\nabla_{\widehat{\mathbf{z}}}D_{KL}\left(q_{\boldsymbol{\theta}}(\mathbf{y}|\mathbf{z})\middle\|q_{\boldsymbol{\theta}}(\mathbf{y}|\widehat{\mathbf{z}})\right)\right|_{\widehat{\mathbf{z}}=\mathbf{z}}\nonumber \\
    &=-\sum_{\mathbf{y}\in\mathcal{Y}}q_{\boldsymbol{\theta}}(\mathbf y|\mathbf{z})\nabla_{\mathbf{\widehat z}}\log q_{\boldsymbol{\theta}}(\mathbf y|\mathbf{\widehat z})\big|_{\widehat{\mathbf{z}}=\mathbf{z}}  \\
    &=-\sum_{\mathbf{y}\in\mathcal{Y}} q_{\boldsymbol{\theta}}(\mathbf y|\mathbf{z}) \frac{\nabla_{\mathbf{z}}q_{\boldsymbol{\theta}}(\mathbf y|\mathbf{z})}{q_{\boldsymbol{\theta}}(\mathbf y|\mathbf{z})}\\
    &=\mathbf{0},\label{eq:first_order_term_4}
    \end{align}
    where~\eqref{eq:first_order_term_4} holds due to the linearity of the gradient operator.
    
    The second order term is computed as follows:
    \begin{align}
        &\left .\nabla_{\widehat{\mathbf{z}}}^2 D_{KL}\left(q_{\boldsymbol{\theta}}(\mathbf{y}|\mathbf{z})\middle\|q_{\boldsymbol{\theta}}(\mathbf{y}|\widehat{\mathbf{z}})\right)\right|_{\widehat{\mathbf{z}}=\mathbf{z}}\nonumber\\
        &=-\sum_{\mathbf{y}\in\mathcal{Y}} q_{\boldsymbol{\theta}}(\mathbf y|\mathbf{z})\nabla^2_{\widehat{\mathbf{z}}}\log q_{\boldsymbol{\theta}}(\mathbf y|\widehat{\mathbf{z}})\big|_{\widehat{\mathbf{z}}=\mathbf{z}}\label{eq:second_order_term2}\\
        &=-\sum_{\mathbf{y}\in\mathcal{Y}} q_{\boldsymbol{\theta}}(\mathbf y|\mathbf{z})\bigg\{
        \frac{\nabla_{\mathbf{z}}^2 q_{\boldsymbol{\theta}}(\mathbf y|\mathbf{z})}{q_{\boldsymbol{\theta}}(\mathbf y|\mathbf{z})}\nonumber\\
        &\mathrel{\phantom{=}}-\nabla_{\mathbf{z}}\log q_{\boldsymbol{\theta}}(\mathbf y|\mathbf{z})\nabla_{\mathbf{z}}\log q_{\boldsymbol{\theta}}(\mathbf y|\mathbf{z})^\mathsf{T}
        \bigg\}\\
        &=-\sum_{\mathbf{y}\in\mathcal{Y}} \bigg\{\nabla_{\mathbf{z}}^2 q_{\boldsymbol{\theta}}(\mathbf y|\mathbf{z})\nonumber\\
        &\mathrel{\phantom{=}} - q_{\boldsymbol{\theta}}(\mathbf y|\mathbf{z}) \nabla_{\mathbf{z}}\log q_{\boldsymbol{\theta}}(\mathbf y|\mathbf{z})\nabla_{\mathbf{z}}\log q_{\boldsymbol{\theta}}(\mathbf y|\mathbf{z})^\mathsf{T}\bigg\}\\
        &=\mathcal{I}(\mathbf{z}),\label{eq:second_order_term6}
    \end{align}
    where \eqref{eq:second_order_term6} follows from $\sum_{\mathbf{y}\in\mathcal{Y}} {\nabla_{\mathbf{z}}^2 q_{\boldsymbol{\theta}}(\mathbf y|\mathbf{z})} = 0$, which holds due to the linearity of the Hessian operator. 
    The Fisher information matrix $\mathcal{I}(\mathbf{z})$ is defined as 
    \begin{equation}
        \mathcal{I}(\mathbf{z})=\mathbb{E}_{q_{\boldsymbol{\theta}}(\mathbf y|\mathbf{z})}\left[\nabla_{\mathbf z} \log q_{\boldsymbol{\theta}}(\mathbf y|\mathbf{z})\nabla_{\mathbf z} \log q_{\boldsymbol{\theta}}(\mathbf y|\mathbf{z})^{\mathsf T}\right].
    \end{equation}
    
    Consequently, the second order Taylor approximation of the KL divergence is given by
    \begin{equation} \label{eq:Taylor_fin}
    D_{KL}\left(q_{\boldsymbol{\theta}}(\mathbf{y}|\mathbf{z})\middle\|q_{\boldsymbol{\theta}}(\mathbf{y}|\widehat{\mathbf{z}})\right)
    \approx \frac{1}{2}(\widehat{\mathbf{z}}-\mathbf{z})^{\mathsf T}\mathcal{I}(\mathbf{z})(\widehat{\mathbf{z}}-\mathbf{z}).
    \end{equation}
    
    \section{Proof of Proposition 2}\label{sec:proof2}
    The conditional expectation of the KL divergence can be approximated as
    \begin{align}\label{eq:expect_n}&\mathbb{E}_{p(\widehat{\mathbf z}|\mathbf z)}\left[ D_{KL}\left(q_{\boldsymbol{\theta}}(\mathbf{y}|\mathbf{z})\middle\|q_{\boldsymbol{\theta}}(\mathbf{y}|\widehat{\mathbf{z}})\right)\right]\nonumber \\
    &\approx \frac{1}{2}\mathbb{E}_{p(\widehat{\mathbf z}|\mathbf z)}\left[(\widehat{\mathbf{z}}-\mathbf{z})^\mathsf{T}\mathcal{I}(\mathbf{z})(\widehat{\mathbf{z}}-\mathbf{z})\right]
    \\&= \frac{1}{2}\mathbb{E}_{p(\mathbf{n})}\left[\operatorname{Tr}\left(\mathbf{n}^\mathsf{T}\mathcal{I}(\mathbf{z})\mathbf{n}\right)\right]\\&= \frac{1}{2}\mathbb{E}_{p(\mathbf{n})}\left[\operatorname{Tr}\left(\mathcal{I}(\mathbf{z})\mathbf{n}\mathbf{n}^\mathsf{T}\right)\right]\\&=\frac{1}{2}\operatorname{Tr}\left(\mathcal{I}(\mathbf{z})\Sigma_\mathbf{n}\right).
    \end{align}
    
    For an AWGN channel with a noise covariance matrix with $\sigma^2I$, the conditional expectation is given by $\frac{\sigma^2}{2}\operatorname{Tr}\left(\mathcal{I}(\mathbf{z})\right)$. 
    For a slow fading channel with a given $h$, the noise covariance matrix is $({\sigma^2}/{|h|^2})\mathbf I$, and the conditional expectation of the KL divergence is $({\sigma^2}/{2|h|^2})\operatorname{Tr}(\mathcal{I}(\mathbf{z}))$.
    Since both regularization terms are proportional to $\operatorname{Tr}(\mathcal{I}(\mathbf{z}))$, they can be incorporated into the loss function as $\lambda \operatorname{Tr}(\mathcal{I}(\mathbf{z}))$.

    \bibliographystyle{IEEEtran}
    \bibliography{abrv,mybib}

\end{document}